\documentclass[11pt]{article}
\usepackage{epsf,amsmath,amssymb}
\newcommand{\code}[1]{{\tt #1}}
\newcommand{\abbrev}{\small}
\newcommand{\ep}{\epsilon}
\newcommand{\eqn}[1]{Eq.\,(\ref{#1})}
\newcommand{\fig}[1]{Fig.\,\ref{#1}}
\newcommand{\tab}[1]{Tab.\,\ref{#1}}
\newcommand{\sct}[1]{Sect.\,\ref{#1}}
\newcommand{\app}[1]{App.\,\ref{#1}}
\newcommand{\reference}[1]{Ref.\,\cite{#1}}

\newcommand{\order}[1]{{\cal O}(#1)}
\newcommand{\bld}[1]{\boldmath{$#1$}}

\renewcommand{\Im}{{\rm Im}}
\newcommand{\rhadf}{\code{rhad}}%
\newcommand{\fortran}{\code{fortran}}
\newcommand{\msbar}{\overline{\mbox{\abbrev MS}}}
\newcommand{\mqbar}[1]{\bar{m}_{#1}}
\newcommand{\cf}{C_{\rm F}}
\newcommand{\ca}{C_{\rm A}}
\newcommand{\tr}{T}
\newcommand{\vcharge}{{\cal V}}

\newcommand{\lMs}{L_{ms}}
\newcommand{\lnsmu}{L_{s\mu}}
\newcommand{\lnmsms}{L_{ms}}
\newcommand{\lmumms}{l_{\mu m}}

\begin{document}

\title{\vskip-2cm{\baselineskip14pt
  \centerline{\hfill\normalsize CERN-TH/2002-379}
  \centerline{\hfill\normalsize DESY 02-223}
  \centerline{\hfill\normalsize hep-ph/0212294}
  \centerline{\hfill\normalsize December 2002}}
\vskip1.5cm
\rhadf{}: a program for the evaluation of the hadronic
$R$-ratio in the perturbative regime of QCD}
\author{Robert~V.~Harlander$^{a}$ and Matthias~Steinhauser$^{b}$
  \\[3em]
  {\normalsize (a) Theory Division, CERN, CH-1211 Geneva 23, 
    Switzerland}
  \\[.5em]
  {\normalsize (b) II. Institut f\"ur Theoretische Physik,}\\ 
  {\normalsize Universit\"at Hamburg, D-22761 Hamburg, Germany}
}
\date{}
\maketitle

\begin{abstract}
  This paper describes the \fortran{} program \rhadf{} which performs
  a numerical evaluation of the photon-induced
  hadronic $R$-ratio, $R(s)$, related to
  the cross section for electron-positron annihilation, for a given
  center-of-mass energy $\sqrt{s}$. In \rhadf{} the state-of-the-art
  perturbative corrections to $R(s)$ are implemented and the running and
  decoupling of the strong coupling constant and the quark masses is
  automatically treated consistently.  Several options allow for a
  flexible use of the program.
\end{abstract}

\thispagestyle{empty}


\newpage


\noindent
{\bf PROGRAM SUMMARY}

\begin{itemize}

\item[]{\it Title of program:}
  \rhadf{}

\item[]{\it Available from:}\\
  {\tt http://www.rhad.de/ }
\item[]{\it Computer for which the program is designed and others on which it
    is operable:}
  Any work-station or PC where \fortran{} is running.
\item[]{\it Operating system or monitor under which the program has been
    tested:} 
  Alpha, Linux, Solaris
\item[]{\it No.\ of bytes in distributed program including test data etc.:}
  $168000$

\item[]{\it Distribution format:} 
  ASCII
  
\item[]{\it Keywords:} Hadronic $R$-ratio, perturbative {\abbrev QCD},
  electron-positron annihilation, running and decoupling of $\alpha_s$
  
\item[]{\it Nature of physical problem:} The hadronic $R$-ratio $R(s)$
  is a fundamental quantitiy in high energy physics. It is defined as
  the ratio of the inclusive cross section $\sigma(e^+ e^-\to
  \mbox{hadrons})$ and the point cross section $\sigma_{pt} =
  4\pi\alpha^2/(3s)$. It is well-defined both from the experimental and
  the theoretical side. $R(s)$ belongs to the few physical quantities
  for which high-order perturbative calculations have been performed
  (partial results up to order $\alpha_s^4$ 
  exist!).  Mass effects from real and
  virtual quarks, the evolution of the $\msbar$ parameters, in
  particular in the presence of thresholds, and other subtleties lead to
  fairly complex results in high orders. Thus it is important to provide
  a comprehensive collection of formulas in order to make them available
  to non-experts.

\item[]{\it Method of solution:}
  \rhadf{} is a compilation of all currently available perturbative
  {\abbrev QCD}
  corrections to the quantity $R(s)$. Several options are provided which allow
  for a flexible use. In addition, \rhadf{} contains routines which perform
  the running and decoupling of the strong coupling constant. Thus only the
  center-of-mass energy has to be provided in order to determine $R(s)$.

\item[]{\it Restrictions on the complexity of the problem:}
  The applicability of \rhadf{} is restricted to the perturbative energy
  regions and does not cover the narrow and broad resonances.

\item[]{\it Typical running time:}
  The typical runtime is of the order of fractions of a second.

\end{itemize}

\newpage


\noindent
{\bf LONG WRITE-UP}


\section{General structure of \bld{R(s)}}\label{sec::genstruc}
\begin{figure}[ht]
  \begin{center}
    \begin{tabular}{cc}
      $(a)\qquad\qquad$ &
      $(b)$      \\[.5em]
      \epsfxsize=10em
      \epsffile[135 550 370 680]{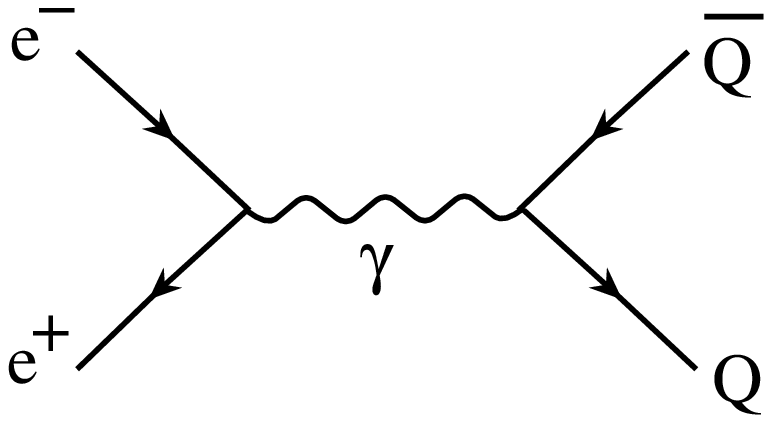}$\qquad\qquad$ &
      \epsfxsize=8em
      \epsffile[150 540 360 680]{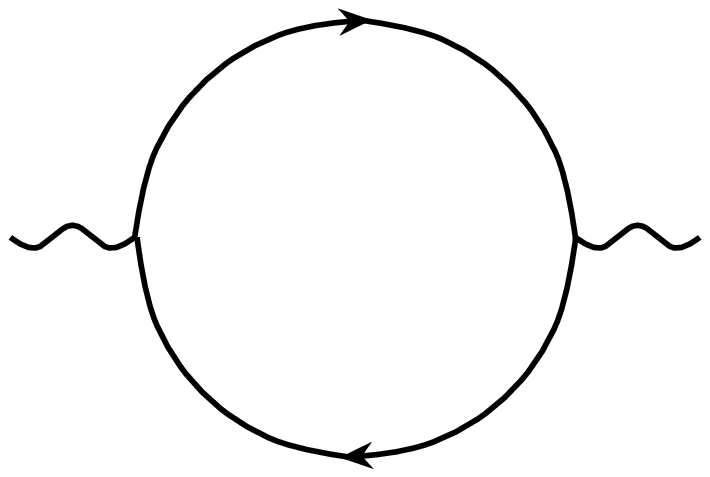}
   \end{tabular}
    \caption[]{\label{fig::tree}
      $(a)$ Tree-level graph for $e^+e^-\to Q\bar Q$. 
      $(b)$ One-loop photon polarization function.
      }
  \end{center}
\end{figure}

We are considering the fully inclusive production of quark pairs in
$e^+e^-$ annihilation (for a review see~\reference{Chetyrkin:1996ia}).  The
tree-level diagram for this process in shown in \fig{fig::tree}\,$(a)$. At
leading order {\abbrev (LO)} perturbative {\abbrev QCD} ({\abbrev
  pQCD}), the cross section as a function of the squared center-of-mass
(c.m.s.) energy, $s$, has thresholds at the points $s=4m_Q^2$ with
$Q=d,u,s,c,b,t$.  However, close to threshold, fixed-order {\abbrev
  pQCD} is no longer applicable.  Below threshold, non-perturbative
effects lead to the formation of bound states of the quark--anti-quark
pair, which appear as more or less sharp peaks in the cross
section.\footnote{Due to the large width no bound state is formed in the
  top quark system.  However, a peak in the cross section remains around
  $s=4m_t^2$.} In our perturbative framework of charm and bottom
production, these narrow resonances cannot be described, and are
therefore not included in \rhadf{}. In addition, we have to spare out
the region between the physical threshold $s_{Q}^{\rm low}$ and the
beginning of the more or less flat continuum region at $s_{Q}^{\rm
  thr}$, where the cross section exhibits rapid variations.  In the case
of charm production, for example, the limits would be $\sqrt{s_{c}^{\rm
    low}} \approx 2m_D \approx 3.73$\,GeV and $\sqrt{s_{c}^{\rm thr}} \approx
4.8$\,GeV, while for bottom production they are $\sqrt{s_{b}^{\rm low}}
\approx 2m_B \approx 10.52$\,GeV and $\sqrt{s_b^{\rm thr}} \approx 11.2$\,GeV.
Even though \rhadf{} provides a numerical value in these threshold
regions, one should not try to give this value any physical
significance; \rhadf{} will print a warning that reminds the user of
this fact.

Furthermore, we have to exclude the low-energy region, below about
$\sqrt{s_{\rm min}} = 1.8$~GeV, where the validity of perturbation
theory is doubtful.  Thus, mass effects from $u$-, $d$- and $s$-quarks
are negligible, which is why we will consider these quarks as massless
throughout the paper.

Instead of the cross section $\sigma(e^+e^-\to {\rm hadrons})$, we will
express the results in terms of the so-called hadronic $R$-ratio,
\begin{equation}
\begin{split}
R(s) = \frac{\sigma(e^+e^-\to {\rm hadrons})}{\sigma(e^+e^-\to
  \mu^+\mu^-)}\,,
\label{eq::rs}
\end{split}
\end{equation}
where
\begin{equation}
\begin{split}
\sigma(e^+e^-\to \mu^+\mu^-) = \frac{4\alpha^2\pi}{3s}
\end{split}
\end{equation}
is the high-energy limit for the photon-mediated
lowest order muon pair production.
In this paper we concentrate on the contributions induced by
photon exchange. $Z$-boson exchange is only relevant at high energies.
There, however, apart from the {\abbrev QCD} corrections\footnote{For a
  discussion of corrections to the axial-vector correlator see
  \reference{Chetyrkin:1996ia}.}, also the electro-weak corrections
become important~\cite{Kuhn:1999kp}; these will not be addressed in this paper.
Nevertheless, \rhadf{} can be used to evaluate the
vector contribution to the production of top quarks in the same way
as for charm and bottom production.

We write $R(s)$ as a sum of contributions from individual quark
flavors, plus a so-called singlet term:
\begin{equation}
\begin{split}
R(s) = \sum_{Q=d,u,s,c,b,t} R_Q(s) + R_{\rm sing}(s)\,.
\end{split}
\end{equation}
The ``non-singlet'' contributions $R_Q(s)$ are all proportional to the
square of the respective quark charge.  The (numerically small) singlet
piece $R_{\rm sing}$ appears for the first time at order $\alpha_s^3$
and will be discussed in \sct{sec::3loop}.

$R_Q(s)$ is defined to be zero below the continuum region of $Q\bar Q$
production ($s_Q^{\rm thr}$), {\it i.e.}, we do not attempt to describe the
resonance regime:
\begin{equation}
\begin{split}
R_Q(s<s_Q^{\rm thr}) = 0\,.
\end{split}
\end{equation}

Our main concern are the radiative corrections to $R_Q(s)$,
predominantly arising from gluons exchanged between, or emitted from,
the $Q\bar Q$ pair, but the lowest order {\abbrev QED} effects, though very
small, will also be included. Thus we write
\begin{equation}
  \begin{split}
    R_{Q}(s) &= 
    \theta(s - s^{\rm thr}_Q)\left[
      \,\sum_{n\geq 0}
      \left(\frac{\alpha_s}{\pi}\right)^n R_{Q}^{(n)}(s)
      + \delta R_{Q}^{\rm QED}(s)
      \right]\,.
  \end{split}
\end{equation}
This formula generally describes the {\abbrev QCD/QED} effects to
$R_Q(s)$, if $\delta R_Q^{\rm QED}(s)$ contains all contributions that
vanish when electro-magnetic corrections are ignored.
Both $\alpha_s$ and $R_Q^{(n)}(s)$ explicitely depend on the
renormalization scale $\mu$, while $R_{Q}(s)$ is invariant under
$\mu$--variation, up to and including the order of perturbation theory
that is being considered.

Since the {\abbrev QCD} corrections only affect the quarks in
\fig{fig::tree}\,$(a)$, the leptonic part of the diagrams can be factored
out and one remains with corrections to the $\gamma Q\bar Q$ vertex, and
real radiation of quarks and gluons. Beyond order $\alpha_s$, the most
convenient way to evaluate the fully inclusive rate is to compute the
photon self energy and relate it to the rate $\gamma^\ast\to Q\bar Q+X$
through the optical theorem, {\it i.e.}, by taking the imaginary part in the
time-like region of the external momentum.
Explicitely,
\begin{equation}
\begin{split}
R(s) = 12\pi\Im\Pi(s+i\ep)\,,
\end{split}
\end{equation}
where
\begin{equation}
\begin{split}
\Pi(q^2) = \frac{1}{3q^2}\left(-g_{\mu\nu} + \frac{q_\mu
    q_\nu}{q^2}\right)\,\Pi_{\mu\nu}(q)
\end{split}
\end{equation}
is the transversal part of the photon polarization function,
$\Pi_{\mu\nu}(q)$.

The different contributions to $R(s)$, as we classify them in our paper,
will be illustrated by sample diagrams contributing to $\Pi(q^2)$ in
what follows. The {\abbrev LO} terms are thus represented by
\fig{fig::tree}\,$(b)$. Since the imaginary part of these diagrams is
obtained by the sum of all possible cuts, it includes real and virtual
contributions simultaneously.

A final remark concerning the counting of loops is in order. We
understand by the $n$-loop {\abbrev QCD} contribution to $R(s)$ the sum
of real and virtual corrections to order $\alpha_s^n$. Computing
$R(s)$ via the optical theorem at $n$-loop level thus requires the
evaluation of the imaginary part of the photon self-energy to $n+1$
loops.


\section{\label{sec::rundec}Running 
  and decoupling of $\alpha_s$ and the quark masses}

As already mentioned in the previous section, the quantity $R(s)$
depends on the strong coupling constant $\alpha_s$, the heavy quark
masses $m_Q$, and the renormalization scale $\mu$.  All of them are
specified in the subroutine {\tt parameters} which is discussed in
Appendix~\ref{app::parameters}.  The dependence of $R(s)$ on $\mu$ is
explicit in terms of logarithms, but also implicit as, for example, in
the argument of $\alpha_s$.  At higher orders in perturbation theory,
one needs to specify the scheme in which $\alpha_s$ and the quark masses
$m_Q$ are evaluated.  For $\alpha_s$, we will always adopt the $\msbar$
scheme which is commonly used in {\abbrev QCD} calculations.  Concerning
the quark masses, it is generally more appropriate to use the pole mass
in the energy region close to the production threshold of the quarks.
On the other hand, the use of the running mass and setting $\mu^2=s$
resums part of the potentially large logarithms $\ln m_Q^2/s$ in the
high-energy region.  For this reason, \rhadf{} provides the logical
parameter {\tt lmsbar} that allows the user to switch between the pole
and the $\msbar$ mass definition for the evaluation of $R(s)$.  If not
stated otherwise, we denote by $m_Q$ a generic quark mass in what
follows.  In those cases where it is necessary to distinguish between
the pole and $\msbar$ quark mass, we denote them by $M_Q$ and $\mqbar{Q}
\equiv \mqbar{Q}^{(n_f)}(\mu)$, respectively. The conversion formula
between $M_Q$ and $\mqbar{Q}$ is given in \app{app::renorm},
\eqn{eq:zminv}.  The ``scale-invariant'' mass is defined recursively as
$\mqbar{Q}(\mqbar{Q}) \equiv \mqbar{Q}^{(n_f)}(\mqbar{Q}^{(n_f)})$,
where the number of active flavors is equal to $n_f = 4$ for charm,
$n_f=5$ for bottom, and $n_f=6$ for top quarks.

If one chooses to evaluate $R(s)$ in terms of $\msbar$ masses, the input
required by \rhadf{} is $\alpha_s^{(5)}(M_Z)$ and the scale invariant
masses $\mqbar{Q}(\mqbar{Q})$ ($Q=c,b,t$). In addition, one needs to
specify the renormalization scale $\mu$ and the c.m.s.\ energy squared,
$s$. From that, \rhadf{} will determine the number of active flavors
$n_f$, as well as the parameters $\alpha_s^{(n_f)}(\mu)$ and
$\mqbar{Q}^{(n_f)}(\mu)$, which will be used for the evaluation of
$R(s)$. As already mentioned before, a natural choice (and the default
value in \rhadf{}) for the renormalization scale is $\mu^2 = s$.

The number of active flavors is determined through the threshold
variables $s_c^{\rm thr}$, $s_b^{\rm thr}$, and $s_t^{\rm thr}$.  When
$s<s_c^{\rm thr}$, it assumes the smallest possible value, $n_f=3$,
and it increases by one every time one of the threshold variables is
crossed, up to $n_f=6$ for $s>s_t^{\rm thr}$.  

On the other hand, in the definition of $\alpha_s^{(n_f)}(\mu)$ one has
the freedom to choose ``matching scales'' $\mu_c$, $\mu_b$, and $\mu_t$,
at which the transition from $n_f$ to $n_f\pm1$ is performed.  For
example, assume that $n_f=4$, and the renormalization scale is set to
some specific value $\mu=\bar \mu$.  The procedure to compute
$\alpha_s^{(4)}(\bar \mu)$ at $n$-loop order as implemented in \rhadf{}
is as follows: In a first step the renormalization group equation
({\abbrev RGE}) for the strong coupling (here and in what follows, we
refer to \app{app::renorm} for the definition of the coefficients
$\beta$, $\zeta_g$, $\gamma_m$, and $\zeta_m$),
\begin{eqnarray}
  \mu^2\frac{{\rm d}}{{\rm d}\mu^2}
  \frac{\alpha_s^{(n_f)}(\mu)}{\pi}
  &=&
  \beta^{(n_f)}\left(\alpha_s^{(n_f)}(\mu)\right)
  \,\,=\,\,
  - \sum_{i=0}^{n-1}
  \beta_i^{(n_f)}\left(\frac{\alpha_s^{(n_f)}(\mu)}{\pi}\right)^{i+2}
  \,,
  \label{eq::beta}
\end{eqnarray}
is solved numerically in order to obtain $\alpha_s^{(5)}(\mu_b)$ from
the input value $\alpha_s^{(5)}(M_Z)$.
The use of the decoupling relation
\begin{eqnarray}
  \alpha_s^{(n_f-1)}(\mu_b) &=&
  \left(\zeta_g\right)^2\,
  \alpha_s^{(n_f)}(\mu_b)
\label{eq::asdec}
\end{eqnarray}
to $(n-1)$-loop order, leads to $\alpha_s^{(4)}(\mu_b)$.  Finally,
applying Eq.~(\ref{eq::beta}) a second time gives the value for
$\alpha_s^{(4)}(\bar\mu)$.  The described procedure is performed
automatically in \rhadf{} using the subroutine {\tt rundecalpha}
(see \app{sec::structure2}).

The evaluation of $\mqbar{Q}^{(n_f)}(\mu)$ from the input $\mqbar{Q}(\mqbar{Q})$
proceeds completely analogously.  The {\abbrev RGE} that governs the
running of the quark masses in the $\msbar$ scheme reads:
\begin{equation}
\begin{split}
  \mu^2\,\frac{d}{d\mu^2}\mqbar{Q}^{(n_f)}(\mu)
  &=
  \mqbar{Q}^{(n_f)}(\mu)\,\gamma_m^{(n_f)}\left(\alpha_s^{(n_f)}(\mu)\right) 
  =-\mqbar{Q}^{(n_f)}(\mu)\,\sum_{i\ge0} \gamma_{m,i}^{(n_f)}
  \left(\frac{\alpha_s^{(n_f)}(\mu)}{\pi}\right)^{i+1}\,.
  \label{eq::gamma}
\end{split}
\end{equation}
Combining Eqs.~(\ref{eq::beta}) and~(\ref{eq::gamma}) leads to
\begin{equation}
\begin{split}
\mqbar{Q}^{(n_f)}(\mu) &=
\mqbar{Q}^{(n_f)}(\mu_0)
\,\frac{c(\alpha_s(\mu)/\pi)}{c(\alpha_s(\mu_0)/\pi)}\,,
\end{split}
\end{equation}
with
\begin{equation}
\begin{split}
c(x) &= x^{c_0}\bigg\{
1 + (c_1 - b_1c_0)\, x 
+ \frac{1}{2}\left[ (c_1 - b_1c_0)^2 + c_2 - b_1c_1 + b_1^2c_0 -
  b_2c_0\right]x^2\\
&+ \bigg[ \frac{1}{6}(c_1 - b_1c_0)^3 + \frac{1}{2}(c_1 - b_1c_0)
\left(c_2 - b_1c_1 + b_1^2c_0 - b_2c_0\right)\\
&+ \frac{1}{3}\left(
  c_3 - b_1c_2 + b_1^2c_1 - b_2c_1 - b_1^3c_0 + 2\,b_1b_2c_0 -
  b_3c_0\right)
\bigg]\,x^3 + \order{x^4}\bigg\}\,,\\
c_i &\equiv \frac{\gamma_{m,i}^{(n_f)}}{\beta_0^{(n_f)}}\,,
\qquad
b_i \equiv \frac{\beta_i^{(n_f)}}{\beta_0^{(n_f)}}\,.
\label{eq::cx}
\end{split}
\end{equation}
If $R(s)$ is evaluated at $n$-loop order, the curly bracket of
\eqn{eq::cx} has to be evaluated up to, and including, the term $\propto
x^{n-1}$. 

The matching between $\mqbar{Q}^{(n_f)}$ and $\mqbar{Q}^{(n_f-1)}$ is determined
through the equation
\begin{equation}
\begin{split}
\mqbar{Q}^{(n_f-1)} = \zeta_m\,\mqbar{Q}^{(n_f)}\,,
\label{eq::mqdec}
\end{split}
\end{equation}
where the function $\zeta_m$ can again be found in
\app{app::renorm}. The procedure for the evaluation of
$\mqbar{Q}^{(n_f)}(\mu)$ from the input $\mqbar{Q}(\mqbar{Q})$ in \rhadf{} is called
{\tt rundecmass} (see \app{sec::structure2}).

Note that the evaluation of $\mqbar{Q}^{(n_f)}(\mu)$ is only required if
the user decides to use $\msbar$ masses for the evaluation of $R(s)$.
In this case, in \code{parameters.f} (see \app{app::parameters}),
\begin{verbatim}
      lmsbar = .true.
\end{verbatim}
and the values for \code{massc}, \code{massb}, \code{masst} have to be
set to the scale invariant masses $\mqbar{c}(\mqbar{c})$,
$\mqbar{b}(\mqbar{b})$, $\mqbar{t}(\mqbar{t})$, respectively.  If one
chooses to use pole masses instead, one defines
\begin{verbatim}
      lmsbar = .false.
\end{verbatim}
and sets \code{massc}, \code{massb}, \code{masst} equal to the on-shell
masses $M_c$, $M_b$, $M_t$, respectively.  \rhadf{} will then simply use
these values throughout the calculation.

There is one exception where we do {\it not} allow a choice between
on-shell and $\msbar$ mass: For ``power-suppressed'' corrections, {\it
  i.e.}, double-bubble diagrams that contain a heavy secondary quark
loop (see \fig{fig::2loop} $(e)$ with $m_2\gg \sqrt{s}, m_1$), the heavy
quark mass is {\it always} inserted in the on-shell scheme.  If the
input for the heavy mass is provided in the $\msbar$ scheme, the
corresponding on-shell value is evaluated within \rhadf{} by the
subroutine {\tt mms2mos} (see \app{sec::structure2}), using the proper
conversion relations~\cite{GraBroGraSch90,CheSte99,MelRit99}.


\section{Notation and tree-level result}\label{sec::tree}
\begin{figure}[ht]
  \begin{center}
    \begin{tabular}{cc}
      $(a)$ &
      $(b)$      \\[.5em]
      \epsfxsize=8em
      \epsffile[150 540 360 680]{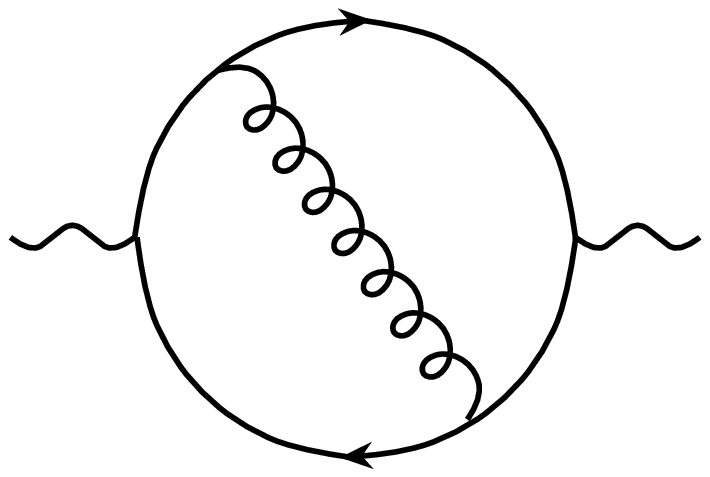} &
      \epsfxsize=8em
      \epsffile[150 540 360 680]{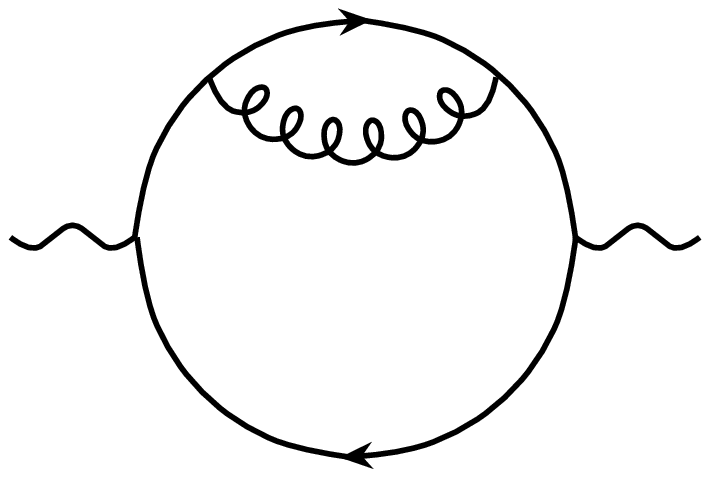} \\
   \end{tabular}
    \caption[]{\label{fig::1loop}
      Diagrams contributing to $R(s)$ at one-loop level.
      }
  \end{center}
\end{figure}
The tree level result for $R_Q(s)$ is part of every
introductory course in quantum field theory. It
reads
\begin{equation}
\begin{split}
R_Q^{(0)}(s) = n_c\,\vcharge_Q^2\,r_V^{(0)}(s,m_Q)\,,\qquad
\mbox{with}\qquad 
r_V^{(0)}(s,M_Q) = \frac{v_Q}{2}(3-v_Q^2)\,,
\label{eq::rv0}
\end{split}
\end{equation}
where $n_c=3$ is the number of colors, and $\vcharge_Q$ is the electric
charge of the quark $Q$ in units of the proton charge. {\it I.e.},
\begin{equation}
\begin{split}
\vcharge_u = \vcharge_c = \vcharge_t = +\frac{2}{3}\,,\qquad
\vcharge_d = \vcharge_s = \vcharge_b = -\frac{1}{3}\,.
\end{split}
\end{equation}
$v_Q$ denotes the quark velocity,
\begin{equation}
\begin{split}
v_Q = \sqrt{1 - \frac{4M_Q^2}{s}}\,,
\label{eq::vq}
\end{split}
\end{equation}
where $\sqrt{s}$ is the c.m.s.\ energy. Note that we define $v_Q$ in
terms of the on-shell mass, $M_Q$ (see \sct{sec::rundec}). This becomes
relevant at higher orders, where we refer to the definition of
\eqn{eq::vq}.


\section{One-loop result}\label{sec::1loop}
Sample diagrams that contribute to the one-loop approximation of $R(s)$
are shown in \fig{fig::1loop}. Their imaginary part has been evaluated a
long time ago in the context of {\abbrev QED}~\cite{Schwinger}. In its
most compact form, the result reads
\begin{equation}
  \begin{split}
    R^{(1)}_Q(s) = n_c \vcharge_Q^2 r_V^{(1)}(s,m_Q)
    \,,
  \end{split}
\end{equation}
with
\begin{equation}
\begin{split}
  r_V^{(1)}(s,M_Q) &= \cf\,\frac{3-v_Q^2}{2}
  \Bigg[
  (1+v_Q^2) \left(
    \mbox{Li}_2\left(p^2\right)
    +2\mbox{Li}_2\left(p\right)
    - \ln p\ln\frac{(1+v_Q)^3}{8v_Q^2}
  \right)
  \\&
  + 3 v_Q \ln\frac{1-v_Q^2}{4v_Q} 
  - v_Q\ln v_Q
  + \frac{33+22v_Q^2-7v_Q^4}{8(3-v_Q^2)}\ln\frac{1+v_Q}{1-v_Q}
  + \frac{15v_Q-9v_Q^3}{4(3-v_Q^2)}
  \Bigg]\,,
  \label{eq::rv1}
\end{split}
\end{equation}
where $p=(1-v_Q)/(1+v_Q)$ and $\cf=4/3$.  In the limit $v_Q\to 1$, one
obtains $r_V^{(1)}=3\cf/4$ which, after taking into account the coupling
and color factors, leads to the following expression for the {\abbrev
  QED} corrections:
\begin{equation}
  \begin{split}
    \delta R_Q^{\rm QED} = n_c \vcharge_Q^4 \frac{3}{4}
    \frac{\alpha}{\pi}\,.
    \label{eq::qed}
  \end{split}
\end{equation}

The tree-level and one-loop functions and their correspondence in
\rhadf{} are summarized in \tab{tab::1loop}.

\begin{table}[ht]
\begin{center}
\renewcommand{\arraystretch}{1.2}
\begin{tabular}{|l|l|c|}
  \hline
  notation & in \rhadf{} & diagrams
  \\\hline  \hline
  $r_V^{(0)}$ & {\tt rv0} & \fig{fig::tree}\,$(b)$ \\\hline
  $r_V^{(1)}$ & {\tt rv1} & \fig{fig::1loop} \\\hline
\end{tabular} 
\end{center}
\caption[]{\label{tab::1loop}
  Contributions to $R(s)$ at tree-level and one-loop order.
  }
\end{table}


\section{Two-loop result}\label{sec::2loop}
\begin{figure}[ht]
  \begin{center}
    \begin{tabular}{ccc}
      $(a)$ &
      $(b)$ &
      $(c)$ \\[.5em]
      \epsfxsize=8em
      \epsffile[150 540 360 680]{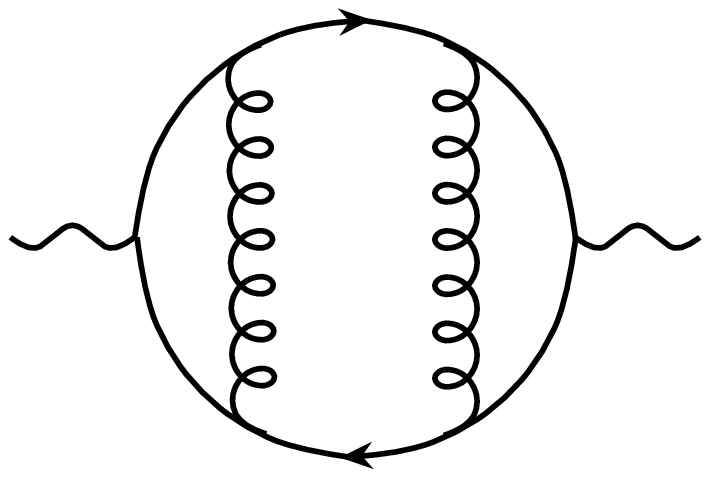} &
      \epsfxsize=8em
      \epsffile[150 540 360 680]{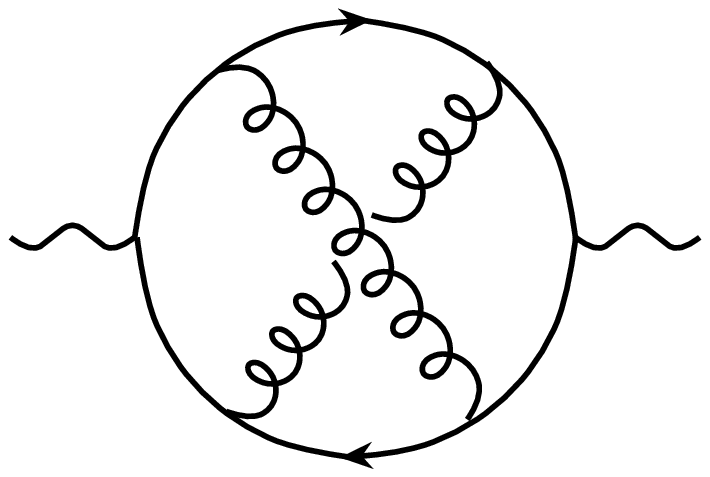} &
      \epsfxsize=8em
      \epsffile[150 540 360 680]{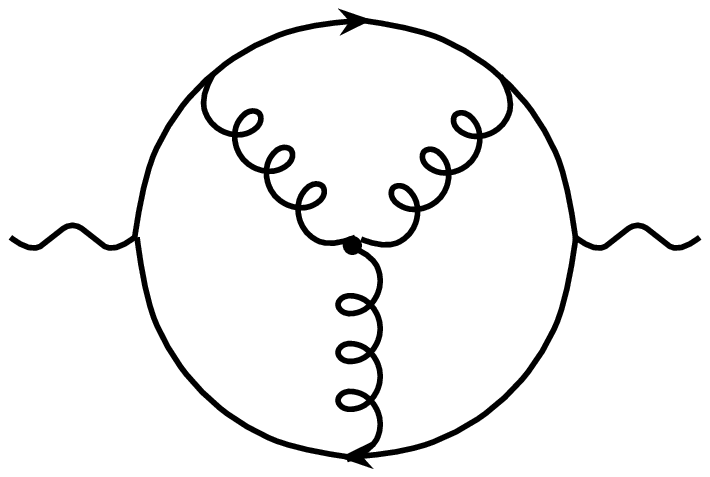}
    \end{tabular}\\
    \begin{tabular}{cc}
      $(d)$ &
      $(e)$ \\[.5em]
    \epsfxsize=8em
      \epsffile[150 540 360 680]{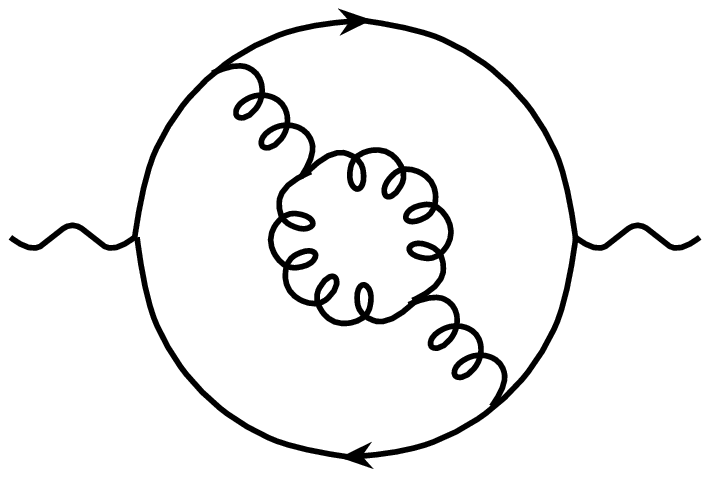} &
      \epsfxsize=8em
      \epsffile[150 540 360 680]{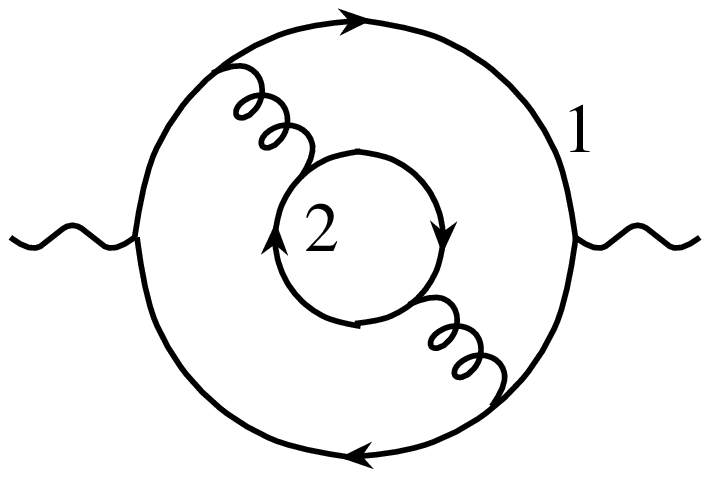}
    \end{tabular}
    \caption[]{\label{fig::2loop}
      Classes of diagrams contributing to $R(s)$ at two-loop level.
      (For the issue of counting loops, see the discussion at the end of
      \sct{sec::genstruc}.)
      }
  \end{center}
\end{figure}

Starting from two-loop order, a general analytic expression for $R(s)$
is no longer available. 
Typical diagrams are shown in \fig{fig::2loop}. The full set
can be obtained from these diagrams 
by attaching the two external photon lines to
the quark lines at arbitrary points.
Only certain classes of diagrams have been
evaluated in closed form. Nevertheless, the analytic evaluation of different
kinematical limits, combined with appropriate interpolation among these
limits, have resulted in extremely accurate semi-analytical
approximations to the full result.
Thus, for all practical purposes, the full mass dependence is available
at order $\alpha_s^2$.

Two main new features occur at the two-loop level that are absent at
lower orders. First, there are contributions that are due to the
non-Abelian character of {\abbrev QCD} and have no correspondence in
{\abbrev QED}. These are diagrams with a three-gluon coupling (see 
\fig{fig::2loop}\,$(c)$ and $(d)$; the four-gluon coupling occurs for
the first time at three-loop order). The second new feature is that a
second quark line can appear, with the effect that the result may depend
on two different quark masses $m_1$ and $m_2$ (see
\fig{fig::2loop}\,$(e)$).

It is clear that the simplest set of diagrams to
evaluate are the ones with self-energy insertions in a gluon propagator,
\fig{fig::2loop}\,$(d)$, $(e)$. In fact, the cases with massless insertions
({\it i.e.}, gluons and massless quarks) are known in closed analytical form,
for general values of $s$ and $m_1$~\cite{HKT,CHKST}. The same is true
for $m_1=0$, $m_2\neq 0$~\cite{HJKT} (labels in accordance with
\fig{fig::2loop}\,$(e)$).  The case $m_1 = m_2$ is known in terms of a
two-dimensional integral representation~\cite{HKT,HT}. For $m_1\ll m_2$,
the coefficient of the leading term in an expansion around $s/m_2^2\to
0$ is known in closed form~\cite{Seidiplom}. For $m_1 = 0$~\cite{Che93}, 
this term was shown to approximate the exact 
$s/m_2^2$ dependence~\cite{HJKT} extremely
well, even up to the threshold $s=4m_2^2$.

The diagrams without self-energy insertions on gluon lines have not yet
been evaluated in closed form for general values of $m_Q$ and $s$. The
massless limit has been known for quite some time~\cite{a2m0}.
Subsequently, mass corrections in the high energy limit have been
computed: The $m_Q^2/s$ terms can be obtained by a simple Taylor
expansion of the integrand~\cite{a2m2}; the $m_Q^4/s^2$ terms were
derived from the $\msbar$ operator product expansion of $\Pi(q^2)$,
combined with renormalization group relations~\cite{CKa2m4}; and the
evaluation of the higher order terms (up to $m_Q^{12}/s^6$) was achieved
by systematically expanding the Feynman integrals~\cite{CHKSm12}.

However, one should note that the convergence of such a high-energy
expansion is not guaranteed below the highest threshold of the diagrams
under consideration. For example, the non-planar diagram in
\fig{fig::2loop}\,$(b)$ has a four-quark cut at $q^2=(4m_Q)^2$, meaning that
the expansion around $m_Q^2/s\to 0$ formally converges only above
$s=(4m_Q)^2$. Nevertheless, in practice one often observes also
satisfactory convergence below the threshold~\cite{CHKSm12}.

A result that is valid in the full kinematic range was obtained in
\reference{CKSpade}
(see also \reference{Steinhauser:2002rq}). 
To this aim, the high-energy limit was combined with 
up to eight moments of the polarization function obtained through an expansion
for $q^2\to0$. This information was combined with 
additional input from the threshold region around $q^2=4 m_Q^2$,
using a conformal mapping and Pad\'e approximation
in order to arrive at a three-loop expression for $\Pi(q^2)$.
For all known practical applications, this approximate
result is equivalent to an analytic expression for $\Pi(q^2)$ and, after
taking the imaginary part, for $R(s)$.

Let us now explicitely parameterize the two-loop contribution to $R(s)$.
We can write
\begin{equation}
\begin{split}
R_{Q}^{(2)}(s) = n_c\,\vcharge_Q^2\,\bigg[&
  r_V^{\rm F}(s,m_Q) + r_V^{\rm A}(s,m_Q) 
  + \sum_q r_V^{\rm db}(s,m_Q,m_q)\bigg]\,,
\end{split}
\label{eq::2lm0}
\end{equation}
where the sum runs over all quark flavors.  $r_V^{\rm F}\propto \cf^2$
denotes contributions from Abelian diagrams, $r_V^{\rm A}\propto \ca\cf$
comes from non-planar and non-Abelian diagrams, and $r_V^{\rm db}\propto
\cf\tr$ (``double-bubble'') arises from diagrams with two separate quark
lines. $\cf = 4/3$, $\ca=3$ and $T=1/2$ are color factors of {\abbrev
  QCD}. For later reference, it is convenient to introduce additional
functions that refer to specific limits of $r_V^{\rm db}$:
\begin{equation}
\begin{split}
r_V^{\rm db}(s,m_Q,0) &\equiv r_V^{\rm L}(s,m_Q)\,,\qquad
r_V^{\rm db}(s,m_Q,m_Q) \equiv r_V^{\rm T}(s,m_Q)\,.
\end{split}
\end{equation}
See also \tab{tab::2loop} for more details.  Note that $r_V^{\rm db}$
introduces contributions from secondary quarks $q$ other than $Q$, both
virtual and real. They arise through the splitting of gluons into a
quark--anti-quark pair.  

In the massless case, one has the simple expressions
\begin{equation}
\begin{split}
r_V^{\rm F}(s,m_Q=0) &= \cf^2\,\left(-\frac{3}{32}\right)\,,\\
r_V^{\rm A}(s,m_Q=0) 
&= \ca\cf\left(
\frac{123}{32} - \frac{11}{4}\,\zeta_3 - \frac{11}{16}\,\lnsmu\right)\,,\\
r_V^{\rm db}(s,m_1=0,m_2=0)
&= \cf\tr\left(-\frac{11}{8} + \zeta_3 + \frac{1}{4}\,\lnsmu\right)\,.
\label{eq::2lml}
\end{split}
\end{equation}
where $\lnsmu \equiv \ln(s/\mu^2)$ and $\zeta_3 \approx 1.20206$. The
expressions in the massive case are given in \app{app::rq2}.

In the following, the expressions adopted for the individual
contributions are described, and their origin is given.

\begin{itemize}
\item For $r_V^{\rm F}(s,m_Q)$ and $r_V^{\rm A}(s,m_Q)$ we use the full
  mass dependence as given by the Pad\'e results of
  Eqs.\,(\ref{eq::rvcf}), (\ref{eq::rvca})~\cite{CKSpade}.
\item For the double-bubble contribution $r_V^{\rm db}(s,m_1,m_2)$, we
  have to distinguish several cases:
  \begin{itemize}
  \item $m_1=m_2 = m_Q$:\quad For $\sqrt{s}\leq 4m_Q$, we use the
    analytic formula as given in~\eqn{eq::r2cc}~\cite{HKT}.  For
    $\sqrt{s}>4m_Q$, the full mass dependence is known in terms of a
    two-fold integral representation, see \eqn{eq::rccr}~\cite{HKT}.  For
    the sake of speed, however, we use the high energy expansion of
    \eqn{eq::rvctm12}~\cite{CHKSm12}. Numerical differences to the
    integral representation are completely negligible.
    Nevertheless, the user may switch to the integral formula by setting
    {\tt lrvctexp = .false.} in \rhadf{}.
  \item $m_2 < m_1$:\quad We apply an expansion for $m_2^2\ll m_1^2 (<
    s/4)$:
    \begin{enumerate}
    \item At $\order{m_2^0}$, the full $m_1$-dependence is used in the
      form of the analytic result of \eqn{eq::rvx} (for
      $x={\rm L}$)~\cite{HKT}.
    \item At $\order{m_2^2}$, we neglect the
      $m_1$ dependence. It turns out that this contribution vanishes 
      which can be seen in \eqn{eq::rvdb0m2}.
    \item Higher orders in $m_2$ are neglected.
    \end{enumerate}
  \item $(2m_1)^2< s < (2m_2)^2$:\quad If $m_1 = 0$, we use the
    analytical result of \eqn{eq::rvdbm2}~\cite{HJKT}, including the
    full dependence on $m_2$.  If $m_1 \neq 0$, we use the leading term
    in $1/m_2^2$, keeping its full $m_1^2$ dependence, see
    \eqn{eq::rvdbts}. This expression has been obtained in
    \reference{Seidiplom}.
  \item $(2m_1)^2 < (2m_2)^2 < s$:\quad If $m_1 = 0$, we use the
    analytical result of \eqn{eq::rvdbm2}~\cite{HJKT}, including the
    full dependence on $m_2$.  If $m_1 \neq 0$, we apply an expansion in
    the limit $m_1^2\ll m_2^2(<s/4)$:
    \begin{enumerate}
    \item At $\order{m_1^0}$, we use the analytical result of
       \eqn{eq::rvdbm2}~\cite{HJKT} for the full $m_2$ dependence.
    \item At $\order{m_1^2}$, we neglect all effects from non-zero
      $m_2$. The corresponding expression for $r_V^{\rm db}(s,m_1,0)$
      is given in Eq.~(\ref{eq::rvdbm1}).
    \item Higher orders in $m_1$ are neglected.
    \end{enumerate}
  \end{itemize}
\end{itemize}
Note that we set the ratios $m_c^2/m_t^2$ and $m_b^2/m_t^2$ to
zero and consider mass corrections due to the presence of an additional
light quark only for $c$-quark effects in $b$-quark production.

For the sake of completeness, let us remark that also the two-loop
corrections of order $\alpha\alpha_s$ and order $\alpha^2$ for massless
quarks~\cite{Kat92} are known.  Numerically, these contributions are
very small.  Nevertheless, we include the mixed corrections of order
$\alpha\alpha_s$ into \rhadf{}, which modifies Eq.~(\ref{eq::qed}) to
\begin{equation}
  \begin{split}
    \delta R_Q^{\rm QED} = n_c \vcharge_Q^4 \frac{3}{4}
    \frac{\alpha}{\pi}
    \left(1 - \frac{1}{3} \frac{\alpha_s(\mu)}{\pi}\right)
    \,.
    \label{eq::qed2}
  \end{split}
\end{equation}
Higher order {\abbrev QED} or mixed {\abbrev QED/QCD} effects are
numerically irrelevant and will be neglected~\cite{surguladze}.

\begin{table}[ht]
\begin{center}
\renewcommand{\arraystretch}{1.2}
\begin{tabular}{|l|l|c|c|c|}
  \hline
  notation & in \rhadf{} & diagrams & $m_1$ & $m_2$ 
  \\\hline  \hline
  $r_V^{\rm F}$ & {\tt rvcf} & $(a)$, $(b)$ & $m_Q$ &
  --  \\\hline
  $r_V^{\rm A}$ & {\tt rvca} & $(b)$, $(c)$, $(d)$ & $m_Q$ &
  --  \\\hline
  $r_V^{{\rm db}}$ & {\tt rvdb} & $(e)$ &
  $m_1$ & $m_2$
  \\\hline
  $r_V^{\rm L}$ 
  & {\tt rvnl} & $(e)$ & $m_Q$ & $0$
  \\\hline
  $r_V^{\rm T}$ & {\tt rvct} & $(e)$ & $m_Q$ & $m_Q$  \\\hline
\end{tabular} 
\end{center}
\caption[]{\label{tab::2loop}
  Different contributions to $R(s)$ at two-loop order.
  The column ``diagrams'' refers to \fig{fig::2loop},
  where a sample diagram is shown for each class.
  }
\end{table}


\section{Three-loop result}\label{sec::3loop}
%

\begin{figure}[ht]
  \begin{center}
    \begin{tabular}{cccc}
      $(a)$ &
      $(b)$ &
      $(c)$ &
      $(d)$ \\[.5em]
      \epsfxsize=8em
      \epsffile[150 540 360 680]{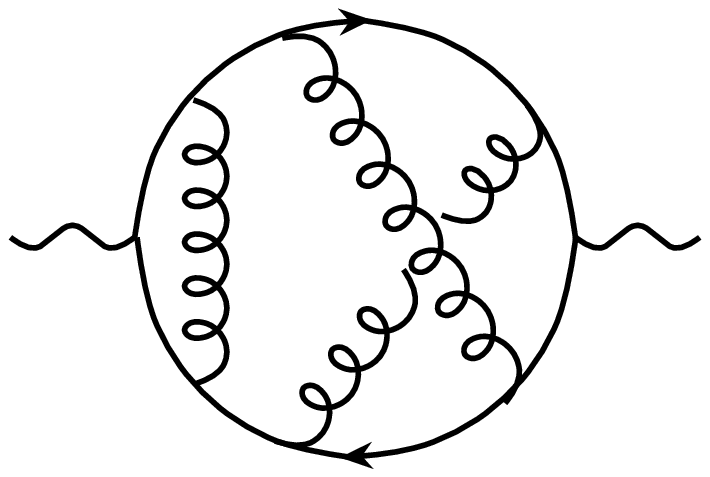} &
      \epsfxsize=8em
      \epsffile[150 540 360 680]{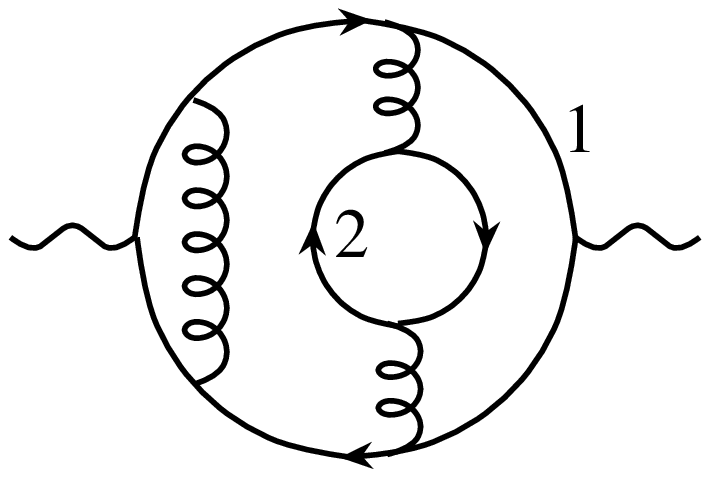} &
      \epsfxsize=8em
      \epsffile[150 540 360 680]{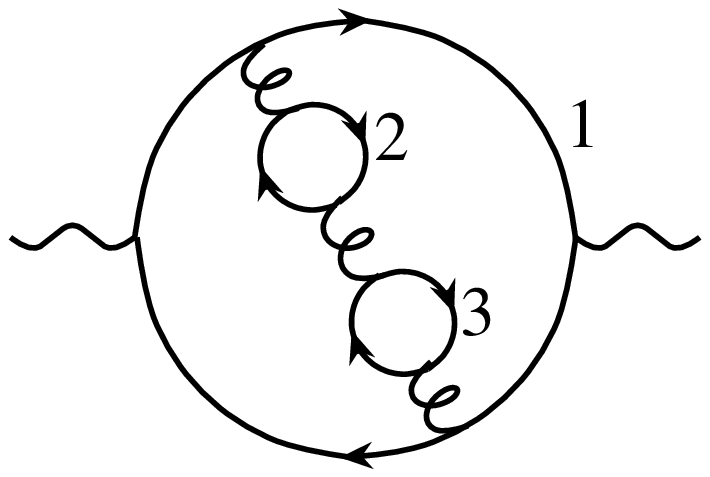} &
      \epsfxsize=9em
      \raisebox{.3em}{
      \epsffile[110 560 315 670]{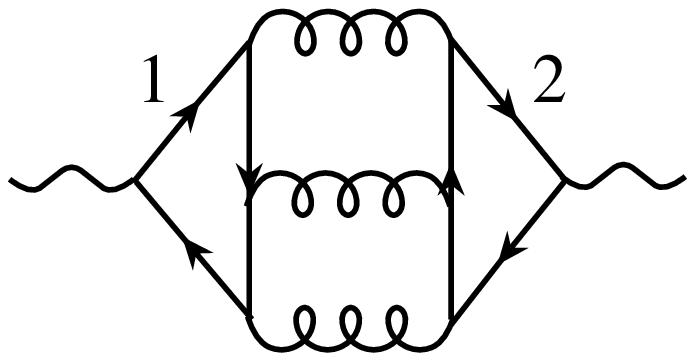}}
  \end{tabular}
  \caption[]{\label{fig::3loop}
      Classes of diagrams contributing to $R(s)$ at three-loop level.
      }
  \end{center}
\end{figure}


\begin{table}
  \begin{center}
    \renewcommand{\arraystretch}{1.2}
    \begin{tabular}{|l|l|c|c|c|c|}
      \hline notation & in \rhadf{} & diagrams & $m_1$ & $m_2$ & $m_3$
      \\
      \hline\hline
      $r_V^{(3)}$ & {\tt rv3} &
      $(a)$, $(b)$, $(c)$  & $m_Q$ & $m_Q$ or 0 & $m_Q$ or 0 \\\hline
      $\delta r_0^{(3)}$ & {\tt delr03} &
      $(b)$, $(c)$  & 0 & $m_Q$ & $m_Q$ or 0 \\\hline
      $r_{V,\rm sing}^{(3)}$ & {\tt rv3sing} &
      $(d)$ & $m_1$ & $m_2$ & --\\\hline
    \end{tabular}
    \caption[]{\label{tab::3loop}
      Different contributions to $R(s)$ at three-loop order (as far as
      available). The column ``diagrams'' refers to \fig{fig::3loop},
      where a sample diagram is shown for each contribution.
      }
  \end{center}
\end{table}


At order $\alpha_s^3$, a new class of diagrams contributes to $R_{\rm
  had}$, the so-called ``singlet diagrams.'' They differ from the
non-singlet diagrams in the sense that, in the singlet case, the
external currents are not connected by a common quark line. A typical
diagram is shown in \fig{fig::3loop}\,$(d)$.  The non-singlet diagrams
(see \fig{fig::3loop}\,$(a)$--$(c)$) are numerically dominant, but
  \rhadf{} includes both contributions. We will describe them in more
detail in what follows.


\subsection{Non-singlet contributions}
The knowledge of $R(s)$ at three-loop level is restricted to the high
energy limit. The massless limit was obtained for the first time in
\reference{3lm0} and later confirmed through an independent calculation
in \reference{chet3l}.

Mass corrections are known up to $\order{m_Q^4/s^2}$: The quadratic
terms were obtained from renormalization group identities~\cite{CKa3m2},
while the quartic mass terms were evaluated by applying the methods of
\reference{CKa2m4} at three-loop level \cite{CHKa3m4}.

Let us briefly discuss how we treat mass effects from massive inner
quark loops. At three-loop level, there can be three different quark
types in one diagram, with, say $m_1 > m_2 > m_3$.  The approximation
that we apply is to take all terms including $m_1^4$ and $m_2^2$,
neglecting mixed terms of order $m_1^2m_2^2$. Mass effects from $m_3$
are also neglected.  In addition, we do not consider power-suppressed
terms, {\it i.e.}\ diagrams that contain a quark $q$ with $s<s_q^{\rm
  thr}$.

With these approximations in mind, we can write the three-loop
contribution as
\begin{equation}
\begin{split}
  R_{Q}^{(3)}(s) &= n_c\,\vcharge_Q^2\,\bigg[ r_V^{(3)}(s,m_Q,n_f) +
  \sum_{q\neq Q}\,\delta r_0^{(3)}(s,m_q,n_f)
  \bigg]\,.
  \label{eq::r31}
\end{split}
\end{equation}
$\delta r_0^{(3)}(s,m_q,n_f)$ denotes the mass corrections due to
massive inner quark loops different from $Q$. According to the
approximations above, it vanishes for $s<s_q^{\rm thr}$ and also for
$m_q=0$.

Note that $n_f$ itself depends on the c.m.s.\ energy.
Assume, for example, that $s<s_c^{\rm thr}$. Then we have $n_f=3$ and
\begin{equation}
\begin{split}
  R_{Q}^{(3)}(s) &= n_c\,\vcharge_{Q}^2\, r_V^{(3)}(s,0,3)\,,\qquad
  Q=u,d,s\,.
  \label{eq::r33}
\end{split}
\end{equation}
For $s_c^{\rm thr} < s < s_b^{\rm thr}$, we have $n_f=4$ and thus the
non-zero contributions are
\begin{equation}
\begin{split}
  R_{Q}^{(3)}(s) &= n_c\,\vcharge_{Q}^2\,\bigg[ r_V^{(3)}(s,0,4) +
  \delta r_0^{(3)}(s,m_c,4)
  \bigg]\,,\qquad Q=u,d,s,\\
\mbox{and}\qquad
R_{c}^{(3)}(s) &= n_c\,\vcharge_c^2\, r_V^{(3)}(s,m_c,4)\,.
  \label{eq::r34}
\end{split}
\end{equation}
For $s_b^{\rm thr} < s < s_t^{\rm thr}$, it is $n_f=5$ and
\begin{equation}
\begin{split}
  R_{Q}^{(3)}(s) &= n_c\,\vcharge_{Q}^2\,\bigg[ r_V^{(3)}(s,0,5) +
  \delta r_0^{(3)}(s,m_c,5) +
  \delta r_0^{(3)}(s,m_b,5)
  \bigg]\,,\qquad Q=u,d,s,\\
  R_{c}^{(3)}(s) &= n_c\,\vcharge_c^2\,\bigg[ r_V^{(3)}(s,m_c,5)
  +\delta r_0^{(3)}(s,m_b,5)\bigg]\,,\\
  R_{b}^{(3)}(s) &= n_c\,\vcharge_b^2\,\bigg[ r_V^{(3)}(s,m_b,5)
  + \delta r_0^{(3)}(s,m_c,5)\bigg]\,.
  \label{eq::r35}
\end{split}
\end{equation}
Finally, for $s>s_t^{\rm thr}$, all masses except for $m_t$ are
neglected, and, with $n_f=6$, we have
\begin{equation}
\begin{split}
  R_{Q}^{(3)}(s) &= n_c\,\vcharge_{Q}^2\,\bigg[ r_V^{(3)}(s,0,6) +
  \delta r_0^{(3)}(s,m_t,6) \bigg]\,,\qquad Q=u,d,c,s,b,\\
  R_{t}^{(3)}(s) &= n_c\,\vcharge_t^2\, r_V^{(3)}(s,m_t,6)\,.
  \label{eq::r36}
\end{split}
\end{equation}
As mentioned before, in \eqn{eq::r35} it is understood that all
$c$-quark mass effects are included only at $\order{m_c^2}$.

The expressions for $r_V^{(3)}(s,m_Q,n_f)$ and $\delta r_0^{(3)}(s,m_Q,n_f)$
are listed in
\app{app::3loop}.


\subsection{Singlet contributions}\label{sec::singlet}
As already mentioned above, singlet diagrams are characterized by the
fact that each of the external photons couples to a separate quark line
(see, {\it e.g.}, \fig{fig::3loop}\,$(d)$).  The singlet contribution arises
first at three-loop level; at two-loop level, where the two quark lines
are connected by only two gluons, it is zero due to Furry's theorem.

We write the singlet contribution to $R(s)$ in the following way:
\begin{equation}
\begin{split}
  R_{\rm sing}(s) &=n_c\,
  \sum_{q,q'}
  \vcharge_{q}\,\vcharge_{q^{\prime}}\,
  r^{(3)}_{V,\rm sing}(s,m_{q},m_{q^{\prime}})\,,
  \label{eq::r3sing}
\end{split}
\end{equation}
where the sum over $q$ and $q^{\prime}$ runs over all active quark
flavors. {\it E.g.}, for $\sqrt{s}=12$~GeV it includes all quarks except for
the top quark.  Again, only the high energy expansion is known. Quadratic mass
terms are absent, and the expression including quartic mass terms reads:
\begin{equation}
\begin{split}
r^{(3)}_{V,\rm sing}(s,m_1,m_2) &=
\frac{55}{216} - \frac{5}{9}\,\zeta_3 
+ \frac{m_1^4}{s^2}\left( -\frac{10}{9} + \frac{25}{3}\,\zeta_3 \right)
+ \frac{m_2^4}{s^2}\left( -\frac{10}{9} + \frac{25}{3}\,\zeta_3 \right)\,.
\label{eq::rv3sing}
\end{split}
\end{equation}
Note that, if $m_1\neq m_2$, we neglect the lighter of the two masses.
Since this is the lowest order where singlet terms occur,
\eqn{eq::rv3sing} is the same in the $\msbar$ and in the
on-shell mass scheme.



\section{Four-loop result}\label{sec::4loop}
\begin{figure}[ht]
  \begin{center}
    \begin{tabular}{cccc}
      $(a)$ &
      $(b)$ &
      $(c)$ &
      $(d)$ \\[.5em]
      \epsfxsize=8em
      \epsffile[150 540 360 680]{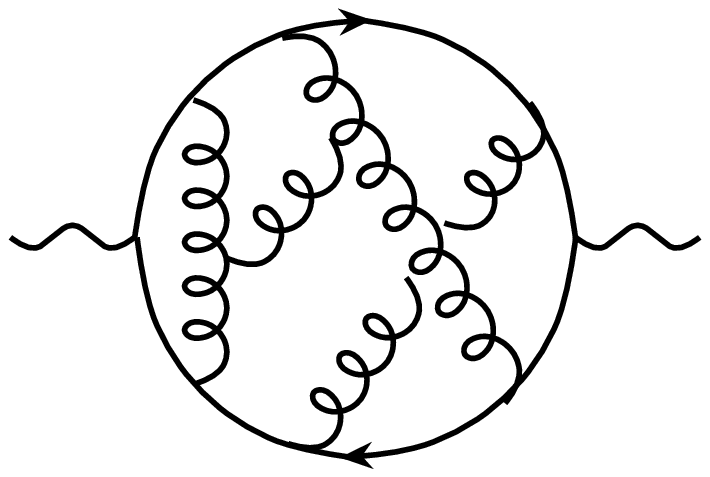} &
      \epsfxsize=8em
      \epsffile[150 540 360 680]{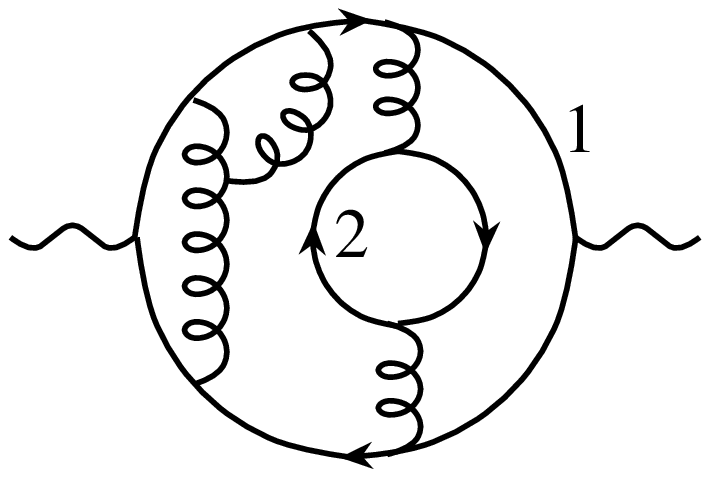} &
      \epsfxsize=8em
      \epsffile[150 540 360 680]{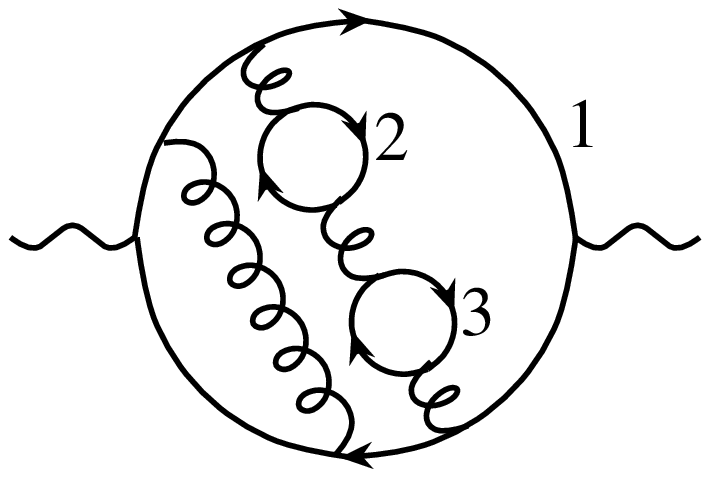} &
      \epsfxsize=8em
      \epsffile[150 540 360 680]{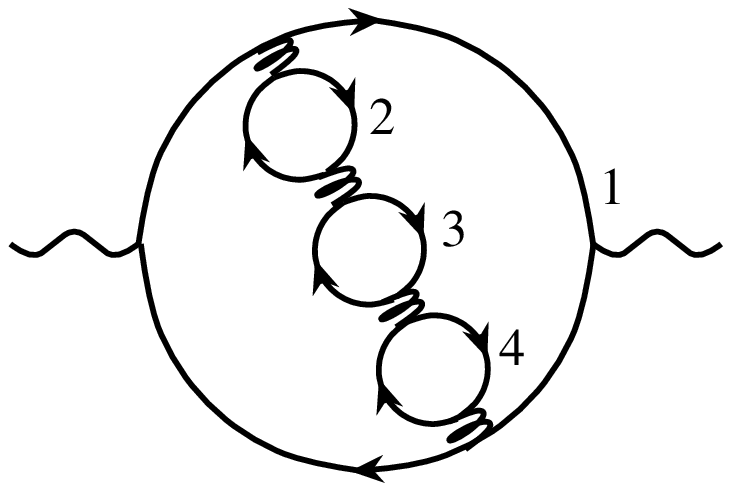} 
    \end{tabular}
    \caption[]{\label{fig::4loop}
      Classes of diagrams contributing to $R(s)$ at four-loop level.
      }
  \end{center}
\end{figure}

The knowledge of the four-loop contribution to $R(s)$ is very limited.
Only contributions of diagrams with two and three quark-loop
insertions have been calculated.
Sample diagrams are shown in
\fig{fig::4loop}\,$(c)$ and $(d)$. The mass effects 
are evaluated up to order $m_Q^2/s$, 
so that the results are strictly only valid in the high-energy limit.


We write the four-loop contribution as
\begin{equation}
\begin{split}
  \bar R_{Q}^{(4)}(s) &= n_c\, \vcharge_Q^2\, \bar{r}_V^{(4)}(s,m_Q)
  \,,
\end{split}
\end{equation}
with
\begin{equation}
\begin{split}
  \bar{r}_V^{(4)}(s,m_Q) &= 
  \bar r_0^{(4),{\rm 0L}}(s) + n_f\,\bar r_0^{(4),{\rm 1L}}(s) 
  + n_f^2\,r_0^{(4),{\rm 2L}}(s) + n_f^3\,r_0^{(4),{\rm 3L}}(s)
  \\&
  + \frac{m_Q^2}{s}\left(
    \bar r_{V,2}^{(4),{\rm 0L}}(s) + n_f\,\bar r_{V,2}^{(4),{\rm 1L}}(s) 
    + n_f^2\,r_{V,2}^{(4),{\rm 2L}}(s) + n_f^3\,r_{V,2}^{(4),{\rm 3L}}(s)
  \right)
  \,,
\end{split}
\end{equation}
where the bar indicates that the expressions are estimates rather than
approximations or even exact results.  $n_f$ is the number of
active flavors, varying with the c.m.s.\ energy as discussed in
\sct{sec::rundec}. 

$r_0^{(4),{\rm 3L}}$ is the renormalon-type contribution with three
massless quark insertions on the gluon propagator, see 
\fig{fig::4loop}\,$(d)$. It has been evaluated in \reference{beneke},
where the general structure of the terms of order $\alpha_s(\alpha_s
n_f)^n$ was derived. Recently, the analytic expression for
$r_0^{(4),{\rm 2L}}$ became available~\cite{BCK}. It required the
evaluation of massless four-loop two-point functions
in combination with the method of 
\reference{chet3l} (see also \reference{Steinhauser:2002rq}) to derive the
imaginary part of the five-loop contributions
as shown in \fig{fig::4loop}\,$(c)$.
The same method has been used in combination with the technique derived
in~\reference{CKa3m2} for the computation of the 
quadratic mass corrections $r_{V,2}^{(4),{\rm 2L}}$ and $r_{V,2}^{(4),{\rm
    3L}}$~\cite{BCKRadcor02}.

Estimates for the full massless four-loop result have been known before
and are still a subject of interest~\cite{KSa4est,a4est,BCKtau}.  We determine
$\bar r_0^{(4)}(s)$ and $\bar r_0^{(4),{\rm L}}(s)$
by subtracting the
known $n_f^2$ and $n_f^3$ contributions from the estimates of
\reference{KSa4est} at
$n_f=1,\ldots,6$, and fitting the resulting six ``data points'' by a
linear function in $n_f$. One finds, for $\mu^2=s$,
\begin{equation}
\begin{split}
\bar r_0^{(4),{\rm 0L}}(s) &= -1.86\cdot 10^{2}\,,\qquad
\bar r_0^{(4),{\rm 1L}}(s) = 21.3\,,\\
r_0^{(4),{\rm 2L}}(s) &= -7.97\cdot 10^{-1}\,,\qquad
r_0^{(4),{\rm 3L}}(s) = 2.15\cdot 10^{-2}\,.
\end{split}
\label{eq::r04}
\end{equation}
The logarithmic contributions follow from the lower order terms through
renormalization group
invariance and are collected in 
App.~\ref{app::4loop}.
Once the exact results for $r_0^{(4),{\rm 0L}}$ and $r_0^{(4),{\rm 1L}}$ become
available, the approximate ones can easily be replaced in \rhadf{}.

Using the same method for the quadratic mass terms in combination
with the estimates of~\reference{BCKRadcor02}
we obtain, in the $\msbar$ scheme
\begin{equation}
\begin{split}
\bar r_{V,2}^{(4),{\rm 0L}}(s) &= 7.11\cdot 10^{3}\,,\qquad
\bar r_{V,2}^{(4),{\rm 1L}}(s) = - 1.43\cdot 10^{3}\,,\\
r_{V,2}^{(4),{\rm 2L}}(s) &= 49.1 \,,\qquad
r_{V,2}^{(4),{\rm 3L}}(s) = - 0.204 \,.
\end{split}
\label{eq::rv4m2}
\end{equation}
In App.~\ref{app::4loop}, the corresponding result for on-shell quark masses is
listed together with the logarithmic contributions.

For completeness, the function implemented in \rhadf{} is listed 
in \tab{tab::4loop}.
Note that at ${\cal O}(\alpha_s^4)$, the corrections analoguous to 
$\delta r_0^{(3)}$ are neglected.

\begin{table}[ht]
\begin{center}
\renewcommand{\arraystretch}{1.2}
\begin{tabular}{|l|l|}
  \hline
  notation & in \rhadf{} \\\hline  \hline
  $r_V^{(4)}$     & {\tt rv4}  \\\hline
\end{tabular} 
\end{center}
\caption[]{\label{tab::4loop}
  Contribution to $R(s)$ at four-loop order.
  }
\end{table}


\section{Evaluating \bld{R(s)}}\label{sec::evalrhad}

In this section we describe how \rhadf{} evaluates $R(s)$ in the individual
energy regions.  Let us first recall that $s_Q^{\rm thr}$ defines the
lowest value of the c.m.s.\ energy squared, $s$, at which the
perturbative treatment of $Q\bar Q$ production is allowed; $R_{Q}(s)$ is
defined to be zero for $s<s_Q^{\rm thr}$. The {\it physical} threshold
for $Q\bar Q$ production is at $s_Q^{\rm low}$; the user is advised,
however, to disregard the results of \rhadf{} in the region between
$s_Q^{\rm low}$ and $s_Q^{\rm thr}$ ($Q=c,b,t$).

The evaluation of the number of active flavors $n_f$, the strong
coupling constant $\alpha_s^{(n_f)}(\mu)$, and, if required, the
$\msbar$ quark masses $\mqbar{Q}^{(n_f)}(\mu)$, from the input values
$s$, $\mu$, $\alpha_s^{(5)}(M_Z)$, and $\mqbar{Q}(\mqbar{Q})$, has been
described in \sct{sec::rundec}. Let us now look in more detail at the
specific contributions that enter $R(s)$ at certain values of $s$.

For $s<s_c^{\rm low}$, {\it i.e.}, below the production threshold for
two $D_0$ mesons, we have $n_f=3$. The charm, bottom and top quark
masses are decoupled and thus their contribution goes like
$\alpha_s^2\,s/m_q^2$ ($q=c,b,t$),
see Eqs.\,(\ref{eq::rvdbm2},\ref{eq::rvdbts}). All other correction terms are
evaluated in the massless limit.

In the region $s_c^{\rm thr}<s<s_b^{\rm low}$ one has $n_f=4$.
$\alpha_s^{(4)}(\mu)$ is evaluated from the input $\alpha_s^{(5)}(M_Z)$,
and, in the $\msbar$ mass scheme, $\mqbar{c}^{(4)}(\mu)$ is obtained
from the input quantity $\mqbar{c}(\mqbar{c})$ (see \sct{sec::rundec}).
In the on-shell mass scheme, the input value $M_c$ is used unchanged
throughout the calculation. The full charm quark mass dependence
(inasmuch it is known) is taken into account. Bottom and top quark are
decoupled and enter through the power-suppressed terms,
Eqs.\,(\ref{eq::rvdbm2},\ref{eq::rvdbts}), at order $\alpha_s^2$.

At c.m.s.\ energies above $s_b^{\rm thr}$ and below $s_t^{\rm low}$, the
situation is similar to the previously described region, but with
$n_f=5$.  Note, however, that in the $\msbar$ scheme not only
$\mqbar{b}^{(5)}(\mu)$ is needed for the numerical evaluation, but also
$\mqbar{c}^{(5)}(\mu)$.  It is obtained from the input
$\mqbar{c}(\mqbar{c})$ using the routine {\tt rundecmass}
(see \sct{sec::structure2}).

For $s>s_t^{\rm thr}$, $\alpha_s^{(6)}(\mu)$ and $\mqbar{t}^{(6)}(\mu)$
are needed for the evaluation of $R(s)$. At these energies, no mass
corrections from lighter quarks are considered.

\tab{tab::2/5GeV} and \ref{tab::12GeV} show the results for $R(s)$
with the default settings (see \sct{app::parameters}), for different
values of the c.m.s.\ energy. The individual rows correspond the
different orders of perturbation theory. In the case of $\sqrt{s} =
5$\,GeV and $\sqrt{s}=12$\,GeV, we also show the partial contributions to
$R(s)$ arising from massive quarks.
The second column contains the value of $\alpha_s^{(n_f)}(\sqrt{s})$,
with the appropriate $n_f$.


\begin{table}
\begin{center}
\begin{tabular}{cc}
\begin{tabular}{|c||c|c|}
\multicolumn{3}{c}{$\sqrt{s} = 2$\,GeV}\\
\hline
 order & $ \alpha_s $ &  $ R(s) $ \\
\hline\hline
 0 &   0.1180 &   2.0012 \\
 \hline 
 1 &   0.2726 &   2.1747 \\
 \hline 
 2 &   0.2981 &   2.2223 \\
 \hline 
 3 &   0.2984 &   2.2049 \\
 \hline 
 4 &   0.2973 &   2.1835 \\
 \hline 
\end{tabular}
&
\begin{tabular}{|c||c|c|c|}
\multicolumn{4}{c}{$\sqrt{s} = 5$\,GeV}\\
\hline
 order & $ \alpha_s $ &  $ R_c(s) $ &
 $ R(s) $ \\
\hline\hline
 0 &   0.1180 &   1.2199 &   3.2220 \\
 \hline 
 1 &   0.2025 &   1.4956 &   3.6267 \\
 \hline 
 2 &   0.2120 &   1.5950 &   3.7467 \\
 \hline 
 3 &   0.2123 &   1.5825 &   3.7272 \\
 \hline 
 4 &   0.2122 &   1.5467 &   3.6867 \\
 \hline 
\end{tabular}
\end{tabular}
\caption[]{\label{tab::2/5GeV} Hadronic $R$-ratio at $\sqrt{s}=2$\,GeV
  ($n_f=3$, left) and $\sqrt{s}=5$\,GeV ($n_f=4$, right).  Note the
  change in $\alpha_s$ which is consistently evaluated from
  $\alpha_s^{(5)}(M_Z)$. The parameter settings are given in
  \app{sec::defpar}.}
\end{center}
\end{table}


\begin{table}
\begin{center}
\begin{tabular}{|c||c|c|c|c|}
\multicolumn{5}{c}{$\sqrt{s} = 12$\,GeV}\\
\hline
 order & $ \alpha_s $ &  $ R_c(s) $ &
 $ R_b(s) $ &
 $ R(s) $ \\
\hline\hline
 0 &   0.1180 &   1.3304 &   0.2675 &   3.6001 \\
 \hline 
 1 &   0.1667 &   1.4199 &   0.3496 &   3.8778 \\
 \hline 
 2 &   0.1706 &   1.4269 &   0.3803 &   3.9265 \\
 \hline 
 3 &   0.1709 &   1.4198 &   0.3797 &   3.9148 \\
 \hline 
 4 &   0.1709 &   1.4158 &   0.3756 &   3.9050 \\
 \hline 
\end{tabular}
\caption[]{\label{tab::12GeV} Hadronic $R$-ratio at $\sqrt{s}=12$\,GeV
  ($n_f=5$).}
\end{center}
\end{table}

\begin{figure}[ht]
  \begin{center}
    \epsfxsize=15cm
    \epsffile[25 230 530 580]{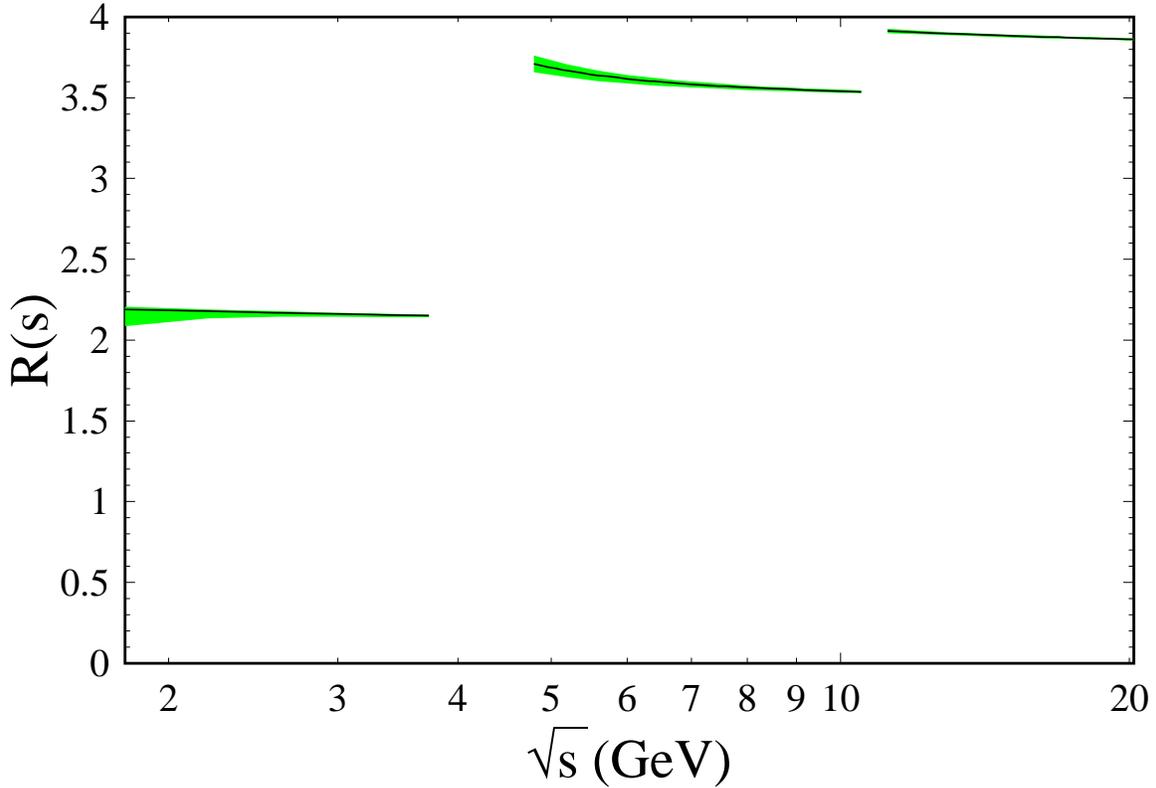}
    \caption[]{\label{fig::rplot}
      $R(s)$ for $1.8$~GeV $<\sqrt{s}< 20$~GeV. The central curve (solid
      line) is obtained using the default values specified in
      App.~\ref{app::parameters} and corresponds to the results given in
      Tab.~\ref{tab::rplot}. The error bands are obtaind by varying
      $\alpha_s(M_Z)$, $\mu$, $M_c$ and $M_b$ within the limits of
      \eqn{eq::limits}.
      Not shown are the perturbatively inaccessible regions between
      $s_Q^{\rm low}$ and $s_Q^{\rm thr}$ ($Q=c,b$, see \sct{sec::genstruc}).
      }
  \end{center}
\end{figure}

Tab.~\ref{tab::rplot} 
contains the results for 
$\alpha_s(s)$, $R(s)$, $R_c(s)$ and $R_b(s)$ for
the energy region between $\sqrt{s}=1.8$~GeV and 
$\sqrt{s}=20$\,GeV, sparing out those parts where
perturbation theory is not applicable. 
We adopted the default values of \rhadf{} given in
App.~\ref{app::parameters} with {\tt iord = 4}. 
In Fig.~\ref{fig::rplot} the data for $R(s)$ (full curve)
are shown in graphical form.
The shaded bands mark the uncertainty of $R(s)$ 
induced by the variation of $\alpha_s(M_Z)$, $\mu$, $M_c$ and $M_b$
in the range
\begin{eqnarray}
  \alpha_s(M_Z) &=& 0.118 \pm 0.003\,,
  \nonumber\\
  M_c &=& (1.65 \pm 0.15)~\mbox{GeV}\,,
  \nonumber\\
  M_b &=& (4.75 \pm 0.2)~\mbox{GeV}\,,
  \nonumber\\
  \mu &\in& \left[\frac{\sqrt{s}}{2},2\sqrt{s}\right] \,.
  \label{eq::limits}
\end{eqnarray}


\begin{table}
\begin{center}
\begin{tabular}{|r|r|r|r|r|}
\hline
$\sqrt{s}$ (GeV) & $ \alpha_s(s) $ &  $ R_c(s) $ &  $ R_b(s) $ &  $ R(s) $ \\
\hline\hline
    1.80 &   0.3148 &   0.0000 &   0.0000 &   2.1895 \\
 \hline 
    2.20 &   0.2833 &   0.0000 &   0.0000 &   2.1780 \\
 \hline 
    2.60 &   0.2620 &   0.0000 &   0.0000 &   2.1684 \\
 \hline 
    3.00 &   0.2463 &   0.0000 &   0.0000 &   2.1607 \\
 \hline 
    3.40 &   0.2342 &   0.0000 &   0.0000 &   2.1544 \\
 \hline 
    3.73 &   0.2260 &   0.0000 &   0.0000 &   2.1499 \\
 \hline 
\hline
    4.80 &   0.2150 &   1.5681 &   0.0000 &   3.7097 \\
 \hline 
    5.00 &   0.2122 &   1.5467 &   0.0000 &   3.6867 \\
 \hline 
    6.00 &   0.2007 &   1.4842 &   0.0000 &   3.6174 \\
 \hline 
    7.00 &   0.1919 &   1.4556 &   0.0000 &   3.5836 \\
 \hline 
    8.00 &   0.1850 &   1.4399 &   0.0000 &   3.5637 \\
 \hline 
    9.00 &   0.1793 &   1.4301 &   0.0000 &   3.5505 \\
 \hline 
   10.00 &   0.1744 &   1.4235 &   0.0000 &   3.5410 \\
 \hline 
   10.52 &   0.1722 &   1.4209 &   0.0000 &   3.5371 \\
 \hline 
\hline
   11.20 &   0.1736 &   1.4186 &   0.3793 &   3.9133 \\
 \hline 
   12.00 &   0.1709 &   1.4158 &   0.3756 &   3.9050 \\
 \hline 
   13.00 &   0.1679 &   1.4129 &   0.3717 &   3.8964 \\
 \hline 
   14.00 &   0.1652 &   1.4106 &   0.3685 &   3.8891 \\
 \hline 
   15.00 &   0.1628 &   1.4087 &   0.3659 &   3.8830 \\
 \hline 
   16.00 &   0.1606 &   1.4070 &   0.3636 &   3.8778 \\
 \hline 
   17.00 &   0.1586 &   1.4056 &   0.3618 &   3.8732 \\
 \hline 
   18.00 &   0.1567 &   1.4044 &   0.3602 &   3.8692 \\
 \hline 
   19.00 &   0.1550 &   1.4033 &   0.3589 &   3.8657 \\
 \hline 
   20.00 &   0.1534 &   1.4023 &   0.3577 &   3.8626 \\
 \hline 
\end{tabular}
\caption[]{\label{tab::rplot}$\alpha_s(s)$, $R_c(s)$,
  $R_b(s)$, and $R(s)$ for various values of $s$.}
\end{center}
\end{table}


\section{\label{sec::test}A typical program}

It is instructive to look at a typical program that evaluates $R(s)$. 
We set the c.m.s.\ energy to $\sqrt{s} = 12$\,GeV, and split
the result according to the contributions from the individual quarks.
The occuring functions will be described in more detail in 
\app{sec::structure}, but their
functionality should be clear from the following program.
\paragraph{Input.}
The corresponding \fortran{} program would look as follows. 
\begin{verbatim}
      program example

      implicit real*8(a-h,m-z)
      implicit integer(i,j)
      implicit character*60(k)
      implicit logical(l)

      include 'common.f'

      sqrts = 12.d0
      scms  = sqrts*sqrts

      call parameters(scms)
      call init(scms)

      rall = rhad(scms)
      ru = ruqrk(scms)
      rd = rdqrk(scms)
      rs = rsqrk(scms)
      rc = rcqrk(scms)
      rb = rbqrk(scms)
      rt = rtqrk(scms)
      rsg = rsinglet(scms)
      rem = rqed(scms)

      print*,'R_had  = ',rall
      print*,'R_u    = ',ru
      print*,'R_d    = ',rd
      print*,'R_s    = ',rs
      print*,'R_c    = ',rc
      print*,'R_b    = ',rb
      print*,'R_t    = ',rt
      print*,'R_sing = ',rsg
      print*,'R_QED  = ',rem

      end
\end{verbatim}

\paragraph{Output.} If {\tt lverbose} is set to {\tt .true.}
the output reads:
\begin{verbatim}

            rhad.f -- Version 1.00
 by Robert Harlander and Matthias Steinhauser
                December 2002

Order of calculation:   4

Scales:
  sqrt(s)  =    12.000 GeV
  mu       =    12.000 GeV
  thrc     =     4.800 GeV
  thrb     =    11.200 GeV
  thrt     =   360.000 GeV

Number of active flavors:
  nf  =  5

Coupling constants:
  alpha_QED    =  1 / ( 137.036 )
  alpha_s(Mz)  =  0.1180             [ Mz =  91.1876 GeV ]
  alpha_s(mu)  =  0.1709360043

Quark masses:
  M_c       =    1.65 GeV
  M_b       =    4.75 GeV
  M_t       =  175.00 GeV

General switches (F=False, T=True):
  only massless terms    : F
  power suppressed terms : T
  QED corrections        : T
  singlet contributions  : T
  alphas^3 m^2 included  : T
  alphas^3 m^4 included  : T
  alphas^4 m^2 included  : T
 -- end of parameters --
 R_had  =   3.9049832
 R_u    =   1.40765446
 R_d    =   0.351913615
 R_s    =   0.351913615
 R_c    =   1.41577367
 R_b    =   0.375554797
 R_t    =   0.
 R_sing =  -4.42977631E-05
 R_QED  =   0.00221733842
\end{verbatim}
If {\tt lverbose = .false.}, only the lines after ``{\tt -- end of
  parameters --}'' are printed.  The parameter list displays the most
important initializations, like the value of $\alpha_s^{(5)}(M_Z)$ and
the renormalization scale $\mu$.  It also contains values for parameters
that were derived from these settings, like the number of active flavors
$n_f$, and the strong coupling constant $\alpha_s^{(n_f)}(\mu)$.


\section{Conclusions}

In this paper we have discussed the perturbative corrections 
to the inclusive cross section $\sigma(e^+e^-\to\mbox{hadrons})$
and collected the analytical and semi-analytical formulas.
This includes the full mass dependence up to order $\alpha_s^2$, the
expansion up to quartic mass terms at order $\alpha_s^3$ and the 
quadratic mass corrections at order $\alpha_s^4$.
Furthermore, the running and decoupling formalism
necessary for a consistent
evaluation of the strong coupling constant and the $\msbar$
quark masses has been presented.

The main subject of this paper, however, is a description of the
\fortran{} program \rhadf{} which allows for the evaluation of $R(s)$,
including all currently available radiative corrections.  The program is
straightforward to use, in particular if the default parameter set is
adopted.  Variations of the physical input values should be done such
that they still resemble the physical case.  The modularity of \rhadf{}
allows for a simple extension once new corrections to the theoretical
predictions for $R(s)$ become available.  In conclusion, \rhadf{} can
easily be used to compute, for example, the perturbative parts of the
hadronic contributions to the running of the electromagnetic coupling,
and the anomalous magnetic moment of the muon.

\vspace{2em}

{\Large\bf Acknowledgments}

First of all, we are indebted to Hans K\"uhn for suggesting this project
and for his valuable advice at all stages of its development.  It is a
pleasure to thank Thomas Teubner for his help and patience in debugging
\rhadf{}.  The comparison to his private code was an essential
ingredient for the completion of this work.  Furthermore, we thank him
and Marek Je\.zabek for providing \fortran{} routines which have been
used in preliminary versions of \rhadf{}.  Thorsten Seidensticker kindly
allowed us to include the unpublished analytical results of $r^{\rm
  db}_V(s,m_1,m_2)$ in the limit $(2m_1)^2< s < (2m_2)^2$.  Furthermore,
we thank the authors of Refs.~\cite{BCKRadcor02,BCKtau} for making the
manuscripts accessible to us before publication.

\vspace{2em}


{\LARGE\bf Appendix}

\begin{appendix}


\section{Installation}

The distribution of \rhadf{} contains the following files:
\begin{verbatim}
Examples  example.f  makefile      r012.f  rhad.f   vegas-rhad.f
common.f  funcs.f    parameters.f  r34.f   runal.f
\end{verbatim}
{\tt Examples} is a directory that contains programs to reproduce
Tables~\ref{tab::2/5GeV}, \ref{tab::12GeV}, and \ref{tab::rplot} of
this paper.  One example program is kept in the main directory,
\code{example.f}.  Its listing in shown in \sct{sec::test}.  It can be
compiled by simply calling {\abbrev GNU} make:
\begin{verbatim}
> gmake prog=example
\end{verbatim}
The executable is named \code{xexample}.  

If {\abbrev GNU} make is not
available, the \fortran{} files can be compiled individually
using 
\begin{verbatim}
f77 -c -o <file>.o <file>.f
\end{verbatim}
and then linked using 
\begin{verbatim}
f77 -o xexample *.o
\end{verbatim}


\section{\label{sec::structure}Basic functions of \rhadf{}}

Unless indicated otherwise, we use the following
implicit type specifications in \rhadf{}:
\begin{verbatim}
      IMPLICIT  REAL*8(A-H,M-Z)
      IMPLICIT  INTEGER(I,J)
      IMPLICIT  CHARACTER*60(K)
      IMPLICIT  LOGICAL(L)
\end{verbatim}

The functions that encode the different loop orders to $R(s)$ 
have already been
discussed in Sects.\,\ref{sec::tree}--\ref{sec::4loop}, and in
particular in Tables\,\ref{tab::1loop}--\ref{tab::4loop}.

The proper sums of these contributions are taken by
the following functions corresponding to 
$R_Q(s)$, $R_{\rm sing}(s)$ and $\delta R_{Q}^{\rm QED}(s)$
(photon exchange only):
\begin{description}
\item[\code{rdqrk(s)}, \code{ruqrk(s)}, \code{rsqrk(s)},
  \code{rcqrk(s)}, \code{rbqrk(s)}, \code{rtqrk(s)}:]\mbox{}\\
  $R_d(s)$,  $R_u(s)$,  $R_s(s)$,  $R_c(s)$,  $R_b(s)$,  $R_t(s)$
\item[\code{rsinglet(s)}:]\mbox{}
  $R_{\rm sing}(s)$
\item[\code{rqed(s)}:]\mbox{}
  $\sum_Q\delta R_{Q}^{\rm QED}(s)$
\end{description}

The full $R$-ratio is obtained by calling \code{rhad(s)}, which adds the
contributions from the individual quarks, depending on the c.m.s.\ 
energy.

The general structure of a \fortran{} program from which the above
functions are called is as follows:
\begin{verbatim}
      program <name of program>
      <declaration of variables>
      <definition of s, e.g.:>
      include 'common.f'
      scms = 11.5d0**2
      call parameters(scms)
      call init(scms)
      <call of function to compute R(s)>
      end
\end{verbatim}
It is important to call the subroutines {\tt parameters} and {\tt init}
(in this order), which define the parameters and evaluate
$\alpha_s(\mu)$ and, if required, evolve and convert the masses.  An
explicit example program has been given in \sct{sec::test}.



\section{\label{app::parameters}The subroutine \code{parameters}}

\subsection{Description of the parameters}

The file \code{parameters.f} contains the subroutine \code{parameters}
and collects the most important variables that can be adjusted by the
user. \code{parameters.f} should be understood as the input file for
\rhadf{}. In Appendix~\ref{sec::defpar}, our default settings are
listed.  Let us stress that, although many of the implemented formulas
are valid over a large parameter range, one should vary the input
parameters only within a sensible region about their default values.
Unreasonable settings, like inverted mass hierarchies or the like, may
lead to inconsistent results.

The following list describes the variables defined in \code{parameters.f}:
\begin{description}
\item[\code{lverbose}:] Print values of parameters (see \sct{sec::evalrhad})
\item[\code{iunit}:] Output unit for parameter list.
\item[\code{iord}:] Order of the calculation. Values:
  \code{iord=0} (Born) to \code{iord=4}.
  It also governs the order at which the running coupling and the
  $\msbar$ masses are evaluated.
\item[\code{alphasmz}:] $\alpha_s^{(5)}(M_Z)$.
  Input value for the strong coupling constant (see \sct{sec::rundec}).
\item[\code{lmsbar}:] If {\tt .true.}, evaluate $R(s)$ by using
  $\msbar$ masses (see \sct{sec::rundec}).
\item[\code{massc, massb, masst}:] Initial values for $m_c$, $m_b$, $m_t$.
  If \code{lmsbar = .true.}: scale invariant mass $\mqbar{q}(\mqbar{q})$; 
  otherwise: on-shell mass $M_q$ (see \sct{sec::rundec}).
\item[\code{mu}:] Renormalization scale $\mu$ at which $\alpha_s$, the
  $\msbar$ quark masses and $R(s)$ are evaluated  (see \sct{sec::rundec}).
\item[\code{muc,mub,mut}:] $\mu_c$, $\mu_b$, $\mu_t$. Matching scales
  for the decoupling of the heavy quarks (see \sct{sec::rundec}).
\item[\code{lqed}:] If {\tt .true.}, include the lowest order {\abbrev QED}
  and, if {\tt iord.ge.2}, the mixed {\abbrev QED/QCD} corrections
  (see Eqs.\,(\ref{eq::qed}) and (\ref{eq::qed2})).
\item[\code{lmassless}:] If \code{.true.}, neglect all quark masses.
\item[\code{lpsup}:] If \code{.true.}, include power-suppressed terms
  (see Eqs.\,(\ref{eq::rhov2}),(\ref{eq::rvdbts})).
\item[\code{la3m2}:] If \code{.true.}, include $\alpha_s^3 m_Q^2$ terms
  (see \sct{app::3loop}).
\item[\code{la3m4}:] If \code{.true.}, include $\alpha_s^3 m_Q^4$ terms
  (see \sct{app::3loop}).
\item[\code{la3sing}:] If \code{.true.}, include singlet terms
  (see \sct{sec::singlet}).
\item[\code{la4m2}:] If \code{.true.}, include $\alpha_s^4 m_Q^2$ terms
  (see \sct{sec::4loop}).
\item[\code{thrc,thrb,thrt}:] $s_c^{\rm thr}$, $s_b^{\rm thr}$, 
  $s_t^{\rm thr}$ (see \sct{sec::genstruc} and \ref{sec::evalrhad}).\\
  Thresholds for open (perturbative) $c$-, $b$-, and $t$-quark pair production.
\item[\code{thrclow,thrblow,thrtlow}:] $s_c^{\rm low}$, $s_b^{\rm low}$,
  $s_t^{\rm low}$ (see \sct{sec::genstruc} and \ref{sec::evalrhad}).\\
  Perturbation theory is not applicable for 
  $s_Q^{\rm low}\le s\le s_Q^{\rm thr}$, $Q\in\{c,b,t\}$.
  Note that these three parameters have no influence on the
  results; their only task is to remind the user of the problematic energy
  regions in the perturbative approach.
\item[\code{sqmin}:] minimal allowed value for the c.m.s.\ energy
  (see \sct{sec::genstruc}).
\end{description}

Note that not all parameters are independent: for example, if {\tt
  lmassless = .true.}, {\tt la3m2}, {\tt la3m4}, and {\tt la4m2} will
  automatically be set to {\tt .false.}.


\subsection{Default settings}\label{sec::defpar}

The default settings in \code{parameters.f} are as follows:
\begin{verbatim}
      subroutine parameters(scms)
c..
c..   User-defined parameters.
c..
      implicit  real*8(a-h,m-z)
      implicit  integer(i,j)
      implicit  character*60(k)
      implicit  logical(l)

      include 'common.f'

c..   verbose mode:
      lverbose = .true.

c..   output unit for parameter list (6 = STDOUT)
      iunit = 6

c..   order or calculation:
      iord  = 4

c..   strong coupling constant at scale mz (5 active flavors):
      alphasmz  = 0.118d0

c..   use MS-bar or pole quark mass?  (.true. == MS-bar mass)
      lmsbar    = .false.

c..   masses
      massc      = 1.65d0        ! charm
      massb      = 4.75d0        ! bottom
      masst      = 175.d0        ! top

c..   renormalization scale:
      mu        = dsqrt(scms)

c..   decoupling scales:
      muc       = 2.d0*massc    ! charm
      mub       = massb         ! bottom
      mut       = masst         ! top

c..   threshold for open quark production
      thrc = 4.8d0
      thrb = 11.2d0
      thrt = 2*masst+10.d0

c..   lower bound of quark threshold region
      thrclow = 3.73d0
      thrblow = 10.52d0
      thrtlow = 2*masst-10.d0

c..   minimum allowed cms energy:
      sqmin     = 1.8d0

c..   some switches
      lmassless = .false.        ! use only massless approximation
      lqed      = .true.         ! QED corrections (.true. = ON)
      lpsup     = .true.         ! include power suppressed terms
      la3m2     = .true.         ! \alpha_s^3 * m^2 terms
      la3m4     = .true.         ! \alpha_s^3 * m^4 terms
      la3sing   = .true.         ! singlet contribution at order \alpha_s^3
      la4m2     = .true.         ! \alpha_s^4 * m^2 terms

      return
      end
\end{verbatim}



\section{\label{sec::structure2}Implementation of running and decoupling}


\paragraph{Determination of \bld{\alpha_s.}}
\rhadf{} contains several routines which cover the consistent
running and decoupling of the strong coupling $\alpha_s^{(n_f)}(\mu)$
(in this context see also Refs.\,\cite{Chetyrkin:2000yt,Steinhauser:2002rq}).
\begin{description}
\item[\code{runalpha(api0,mu0,mu,inf,inloop,verb,apiout):}]\mbox{}\\[-1em]
  \begin{description}
  \item[\code{api0}:] $\alpha_s^{(n_f)}(\mu_0)/\pi$
  \item[\code{mu0}:]  $\mu_0$
  \item[\code{mu}:]   $\mu$
  \item[\code{inf}:]   $n_f$
  \item[\code{inloop}:] number of loops used for {\abbrev QCD} $\beta$ function
  \item[\code{verb}:] 0=quiet,  1=verbose
  \item[\code{apiout}:] $\alpha_s^{(n_f)}(\mu)/\pi$
  \end{description}
  Evaluates $\alpha_s^{(n_f)}(\mu)$ from $\alpha_s^{(n_f)}(\mu_0)$
  by solving numerically the renormalization group equation
\item[\code{decalpha(als,massth,muth,inf,inloop,idir,alsout):}]
  \mbox{}\\[-1em]
  \begin{description}
  \item[\code{inf}:] $n_i$
  \item[\code{massth}:] $M_h$, heavy quark mass to be decoupled
  \item[\code{muth}:] $\mu^{\rm th}$
  \item[\code{inloop}:] $k_{\rm ord}$
  \item[\code{idir}:] $\delta \in \{-1,1\}$
  \item[\code{als}:] $\alpha_s^{(n_i)}$
  \item[\code{alsout}:] $\alpha_s^{(n_i+\delta)}$
  \end{description}
  Evaluates $\alpha_s^{(n_i+\delta)}(\mu^{\rm th})$ from
  $\alpha_s^{(n_i)}(\mu^{\rm th})$.
  The order of the matching relations is determined by $k_{\rm ord}$
  ($0\leq k_{\rm ord} \leq 3$). The matching coefficients require the pole
  quark mass as input.
\item[\code{rundecalpha(als0,mu0,mu1,alsout)}:]\mbox{}\\[-1em]
  \begin{description}
  \item[\code{mu0}:] $\mu_0$
  \item[\code{mu1}:] $\mu_1$
  \item[\code{als0}:] $\alpha_s^{(n_{i})}(\mu_0)$, where $n_i$ is
    defined in the subroutine {\tt parameters}
  \item[\code{alsout}:] $\alpha_s^{(n_{f})}(\mu_1)$, where $n_f$ is
    determined with the help of the variables {\tt thrc}, {\tt thrb} and 
    {\tt thrt} 
  \end{description}
  Determines $\alpha_s^{(n_{f})}(\mu_1)$ from
  $\alpha_s^{(n_{i})}(\mu_0)$, including decoupling and matching, by
  calling \code{decalpha} and \code{runalpha}.
  The value for $n_i=$\code{infini} is set in the subroutines
  \code{parameters}. $n_f=$\code{inffin} is determined according to the 
  value of \code{mu1} and becomes part of the \code{common}-block.
\end{description}
In addition there is the subroutine \code{decalphams} which works
analoguously to \code{decalpha}, except that the matching coefficients are 
parameterized in terms of $\msbar$ quark masses.


\paragraph{Determination of \bld{\mqbar{Q}(\sqrt{s})}.}
\begin{description}
\item[\code{runmass(mass0,api0,apif,inf,inloop,massout)}:]\mbox{}\\[-1em]
  \begin{description}
  \item[\code{mass0}:] $\mqbar{Q}(\mu_0)$
  \item[\code{api0}:] $\alpha_s(\mu_0)/\pi$
  \item[\code{apif}:] $\alpha_s(\mu_f)/\pi$
  \item[\code{inf}:] $n_f$
  \item[\code{inloop}:] number of loops
  \item[\code{massout}:] $\mqbar{Q}(\mu_f)$
  \end{description}
  Evaluates $\mqbar{Q}^{(n_f)}(\mu)$ from $\mqbar{Q}^{(n_f)}(\mu_0)$
  using \eqn{eq::cx}.
\item[\code{decmassms(mq,als,massth,muth,inf,inloop,idir,mqout):}]
  \mbox{}\\[-1em]
  \begin{description}
  \item[\code{mq}:] $\mqbar{Q}^{(n_i)}(\mu^{\rm th})$
  \item[\code{als}:] $\alpha_s^{(n_i)}(\mu^{\rm th})$
  \item[\code{massth}:] $\mqbar{h}$, 
    $\msbar$ value of heavy quark mass to be decoupled 
  \item[\code{muth}:] $\mu^{\rm th}$ decoupling scale
  \item[\code{inf}:] $n_i$
  \item[\code{idir}:] $\delta\in\{-1,1\}$
  \item[\code{inloop}:] $k_{\rm ord}$
  \item[\code{mqout}:] $\mqbar{Q}^{(n_i+\delta)}(\mu^{\rm th})$
  \end{description}  
  Evaluates $\mqbar{Q}^{(n_i+\delta)}(\mu^{\rm th})$ from
  $\mqbar{Q}^{(n_i)}(\mu^{\rm th})$.
  The order of the matching relations is determined by $k_{\rm ord}$
  ($0\leq k_{\rm ord} \leq 3$). The matching coefficients require the 
  $\msbar$ quark mass as input.
\item[\code{rundecmass(mq0,inf,mu0,mu1,mqout)}:]\mbox{}\\[-1em]
  \begin{description}
  \item[\code{mq0}:] $\mqbar{Q}^{(n_{i})}(\mu_0)$
  \item[\code{inf}:] $n_i$
  \item[\code{mu0}:] $\mu_0$
  \item[\code{mu1}:] $\mu_1$
  \item[\code{mqout}:] $\mqbar{Q}^{(n_{f})}(\mu_1)$
  \end{description}  
  Determines the $\msbar$ mass $\mqbar{Q}^{(n_{f})}(\mu_1)$ from
  $\mqbar{Q}^{(n_{i})}(\mu_0)$, including decoupling and matching, by
  calling \code{decmassms} and \code{runmass}.
  $n_f$ is determined according to the 
  value of \code{mu1}.
\end{description}
Note that in \code{decmassms} and \code{rundecmass} only those cases are
implemented which are needed for \rhadf{}.


\paragraph{Pole mass versus $\msbar$ mass.}
In case the $\msbar$ quark masses are used as input 
the numerical values for the pole quark masses, which are needed for certain
(double-bubble) contributions as described in \sct{sec::rundec}, are 
computed with the help of the routine \code{mms2mos}.

\begin{description}
\item[\code{mms2mos(mms,inf,inloop,mos)}:]\mbox{}\\[-1em]
  \begin{description}
  \item[\code{mms}:] $\mqbar{Q}(\mqbar{Q})$ 
  \item[\code{inf}:] number of active flavors
  \item[\code{inloop}:] number of loops used for the conversion
  \item[\code{mos}:] $M_Q$
  \end{description}
  Determines the on-shell mass $M_Q$ from the scale-invariant mass
  $\mqbar{Q}(\mqbar{Q})$, where the \code{inloop}-loop relation is used.
  In practice we use \code{inloop}=\code{iord}.  For the conversion also
  $\alpha_s(\mqbar{Q})$ is needed which we determine with the help
  of the subroutines
  \code{runalpha} and \code{decalphams}.\\
\end{description}



\section{Analytical results}

All results presented in this section are parametrized in terms of
pole quark masses denoted by $M_Q$, $M_1$ or $M_2$.
The conversion to the $\msbar$ formulas can be performed
with the help of Eq.~(\ref{eq:zminv}).

\subsection{\label{app::rq2}Analytical results for $R_Q^{(2)}(s)$}

In this appendix we list the results for the two-loop contributions to
$R(s)$ as they are implemented in \rhadf{}. They are either known
analytically, or in terms of semi-analytical Pad\'e approximants.

The Abelian two-loop contribution reads
\begin{eqnarray}
  r_V^{\rm F}(s,M_Q) &=& 
  r_V^{{\rm F},pa}   - 4\,\cf\,r_V^{(1)}(s,M_Q) \,,
  \label{eq::rvcf}
\end{eqnarray}
where $r_V^{{\rm F},pa}$ is known in the form of Pad\'e
approximants~\cite{CKSpade}. We quote the result based on a
$[5/5]$-Pad\'e approximant:
%

\begin{eqnarray}
  r_V^{{\rm F},pa} &=& \cf^2\,\frac{3}{4\pi}\mbox{Im}\Bigg\{
  \frac{i\pi}{48z}\left[
    (687+186z)\sqrt{1-1/z} 
    + 216\ln\frac{1 - \sqrt{1 - 1/z}}{1 + \sqrt{1 - 1/z}}
  \right]
  +\frac{(1+\,\tilde{\omega})^2}{1-\,\tilde{\omega}}
  \nonumber\\&&\mbox{}
  \times
  \Big[
  2659.1447467995658619093787 + 
  994.90626341783737509012835 \,\tilde{\omega}
  \nonumber\\&&\mbox{}
  \quad
  -
  2742.9905230464882915790412 \,\tilde{\omega}^2 - 
  779.40067776480109262194206 \,\tilde{\omega}^3 
  \nonumber\\&&\mbox{}
  \quad
  +
  562.98408428934139747541229 \,\tilde{\omega}^4 + 
  132.845153672444424204018458\,\tilde{\omega}^5
  \Big]
  \nonumber\\&&\mbox{}
  \times
  \Big[
  121.9768095064564342458129488+ 
  11.5590492918470301261688922 \,\tilde{\omega}
  \nonumber\\&&\mbox{}
  \quad
  -
  119.511603603837393164213484 \,\tilde{\omega}^2 - 
  1.04113850838776740702415214 \,\tilde{\omega}^3 
  \nonumber\\&&\mbox{}
  \quad
  +
  18.4653468020861357442841578 \,\tilde{\omega}^4 + 
  \,\tilde{\omega}^5
  \Big]^{-1} 
  \Bigg\}
  \,,
\end{eqnarray}

%
and
\begin{eqnarray}
  z &=& \frac{1}{1-v_Q^2}\,,\qquad
  v_Q = \sqrt{1 - \frac{4M_Q^2}{s}}
  \,,\qquad
  \tilde{\omega} = \frac{2}{z}-1+2i\sqrt{\frac{1}{z}-\frac{1}{z^2}}
  \,.
\end{eqnarray}
$r_V^{(1)}$ can be found in Eq.~(\ref{eq::rv1}).

The non-Abelian contribution can be written in the
form~\cite{CKSpade}
\begin{eqnarray}
  r_V^{\rm A}(s,M_Q) &=& r_V^{{\rm A},pa} 
  + r_V^{{\rm A},g}(s,M_Q)\Big|_{\xi=4}\,,
  \label{eq::rvca}
\end{eqnarray}
where $\xi$ is the {\abbrev QCD} gauge parameter, $r_V^{{\rm A},g}$ is the
result from the gluonic double-bubble diagram given below and
$r_V^{{\rm A},pa}$ is again given in terms of a Pad\'e approximant:
%

\begin{eqnarray}
  r_V^{{\rm A},pa} &=& \ca\cf\,\frac{3}{4\pi}\mbox{Im}\Bigg\{
  - i\pi\frac{33 - 27 z - 6 z^2}{8z^2\sqrt{1-1/z}}
  + ( 1 + \tilde{\omega})^2 
  \nonumber\\&&\mbox{}
  \times
  \Big[53.56935816521245 
  + 23.42800818317322  \,\tilde{\omega}
  - 28.58891755403711  \,\tilde{\omega}^2
  \nonumber\\&&\mbox{}
  \quad
  - 10.11103466811403  \,\tilde{\omega}^3
  + 0.4266396586616486 \,\tilde{\omega}^4
  \Big]
  \nonumber\\&&\mbox{}
  \times
  \Big[ 19.19224589510177 
  - 1.767625499084376 \,\tilde{\omega} 
  - 13.2990319119615  \,\tilde{\omega}^2 
  \nonumber\\&&\mbox{}
  \quad
  + 1.00664474828887  \,\tilde{\omega}^3 
  +                   \,\tilde{\omega}^4
  \Big]^{-1}
  \Bigg\}
  \,.
\end{eqnarray}

%

Following \reference{Chetyrkin:1996yp}, the exact results
of the imaginary part of the fermionic and gluonic double-bubble 
diagram (see \fig{fig::2loop}\,$(d)$ and $(e)$),
$r_V^{\rm L}$ and $r_V^{{\rm A},g}$, respectively, may be written as
($x\in\{{\rm L;A},g\}$)
\begin{eqnarray}
  r_V^x(s,M_Q) &=& C_x\left[-\frac{1}{3}\left(
      R_\infty^x \ln\frac{\mu^2}{s} - R^x_0
    \right)\frac{r_V^{(1)}(s,M_Q)}{\cf}
    + R_\infty^x \delta^{(2)}
    \right]\,,
    \label{eq::rvx}
\end{eqnarray}
with
\begin{eqnarray}
   C_{\rm L} &=& \cf \tr\,, 
  \nonumber\\
   C_{{\rm A},g} &=& \ca\cf\,, 
  \nonumber\\
   R_\infty^{\rm L}  &=&   1\,,
  \nonumber\\
   R_0^{\rm L}  &=&  - \frac{5}{3}+\ln 4 \,,
  \nonumber\\
   R_\infty^{{\rm A},g} &=&  -\frac{5}{4} - \frac{3}{8}\xi\,,
  \nonumber\\
   R_0^{{\rm A},g} &=& \frac{31}{12}-\frac{3}{4}\xi+\frac{3}{16}\xi^2
  +\left(-\frac{5}{4}-\frac{3}{8}\xi\right)\ln 4
  \,.
\end{eqnarray}
$r_V^{(1)}$ is defined in~(\ref{eq::rv1}),
whereas $\delta^{(2)}$ was originally calculated in \reference{HKT}
and reads:

\begin{eqnarray}
\lefteqn{
\delta^{(2)}  = 
- \frac{\left( 3 - {{v_Q }^2} \right) 
       \left( 1 + {{v_Q }^2} \right) }{6}}\nonumber\\ 
 & & \mbox{}
     \times\bigg\{\mbox{Li}_3(p) - 2\mbox{Li}_3(1 - p) - 
       3\mbox{Li}_3({p^2}) - 4\mbox{Li}_3\Big({p\over {1 + p}}\Big) - 
       5\mbox{Li}_3(1 - {p^2}) + 
       \frac{11}{2}\zeta_3\nonumber\\ 
 & & \mbox{}\quad + 
       \mbox{Li}_2(p)\ln\Big(\frac{4\left( 1 - {{v_Q }^2} \right) }{
          {{v_Q }^4}}\Big) + 2\mbox{Li}_2({p^2})
        \ln\Big(\frac{1 - {{v_Q }^2}}{2{{v_Q }^2}}\Big) + 
       2\zeta_2\bigg[ \ln p - 
          \ln\Big(\frac{1 - {{v_Q }^2}}{4v_Q }\Big) \bigg] 
      \nonumber
        \\ 
 & & \mbox{}\quad - 
       \frac{1}{6}\ln\Big(\frac{1 + v_Q }{2}\Big)
        \bigg[ 36\ln 2\ln p - 44\ln^2 p + 
          49\ln p\ln\Big(\frac{1 - {{v_Q }^2}}{4}\Big) + 
          \ln^2\Big(\frac{1 - {{v_Q }^2}}{4}\Big) \bigg] \nonumber
        \\ 
 & & \mbox{}\quad - 
       \frac{1}{2}\ln p\ln v_Q
        \bigg[ 36\ln 2 + 21\ln p + 16\ln v_Q  - 
          22\ln(1 - {{v_Q }^2}) \bigg]  \bigg\} \nonumber
     \\ 
 & & \mbox{}   + 
  \frac{1}{24}\bigg\{ 
      ( 15 - 6{{v_Q }^2} - {{v_Q }^4} ) 
      \Big( \mbox{Li}_2(p) + \mbox{Li}_2({p^2}) \Big)  + 
     3( 7 - 22{{v_Q }^2} + 7{{v_Q }^4} ) 
      \mbox{Li}_2(p)\nonumber\\ 
 & & \mbox{}\quad - 
     ( 1 - v_Q  ) 
      ( 51 - 45v_Q  - 27{{v_Q }^2} + 5{{v_Q }^3} ) 
      \zeta_2\nonumber\\[2mm] 
 & & \mbox{}\quad + 
     \frac{\left( 1 + v_Q  \right) 
        \left( -9 + 33v_Q  - 9{{v_Q }^2} - 15{{v_Q }^3} + 
          4{{v_Q }^4} \right) }{v_Q }\ln^2 p\nonumber
      \\ 
 & & \mbox{}\quad + 
     \bigg[ ( 33 + 22{{v_Q }^2} - 7{{v_Q }^4} ) 
         \ln 2 - 10( 3 - {{v_Q }^2} ) 
         ( 1 + {{v_Q }^2} ) \ln v_Q 
   \nonumber    \\ 
 & & \mbox{}\quad\quad\quad   - 
        ( 15 - 22{{v_Q }^2} + 3{{v_Q }^4} ) 
         \ln\Big(\frac{1 - {{v_Q }^2}}{4{{v_Q }^2}}\Big)\bigg] 
      \ln p\nonumber\\ 
 & & \mbox{}\quad + 
     2v_Q ( 3 - {{v_Q }^2} ) 
   \ln\Big(\frac{4\left( 1 - {{v_Q }^2} \right) }{{{v_Q }^4}}\Big)
   \bigg[ \ln v_Q - 3\ln\Big(\frac{1 - {{v_Q }^2}}{4v_Q }\Big) 
      \bigg]
       \nonumber\\ 
 & & \mbox{}\quad + 
     \frac{237 - 96v_Q  + 62{{v_Q }^2} + 32{{v_Q }^3} - 
        59{{v_Q }^4}}{4}\ln p - 
     16v_Q ( 3 - {{v_Q }^2} ) \ln\Big(\frac{1 + v_Q }{4}\Big)
      \nonumber\\ 
 & & \mbox{}\quad - 
     2v_Q ( 39 - 17{{v_Q }^2} ) 
      \ln\Big(\frac{1 - {{v_Q }^2}}{2{{v_Q }^2}}\Big) - 
     \frac{v_Q \left( 75 - 29{{v_Q }^2} \right) }{2}\bigg\} 
\,,
\end{eqnarray}

with $\zeta_2= \pi^2/6\approx 1.64493$, $\zeta_3\approx 1.20206$, and
\begin{eqnarray}
  p &=& \frac{1-v_Q}{1+v_Q}\,.
  \nonumber
\end{eqnarray}

The double-bubble contribution $r_V^{\rm T}$ can be written in the form
\begin{eqnarray}
  r_V^{\rm T}(s,M_Q) &=&  T \cf\,\left(
    \rho^V(M_Q^2,M_Q^2,s) + \rho^R(M_Q^2,M_Q^2,s) + 
    \frac{1}{3} \ln\frac{M_Q^2}{\mu^2} \frac{r_V^{(1)}}{3\cf} \right)
  \,,
\label{eq::r2cc}
\end{eqnarray}
where
\begin{eqnarray}
\rho^V(M_Q^2,M_Q^2,s) &=& \frac{1}{6} \,\Bigg[\, 
     \frac{3+10v_Q^2-5v_Q^4}{24} \,\ln^3 p +
     \frac{-3+40v_Q^2+16v_Q^4-15v_Q^6}{12v_Q^3} \,\ln^2 p  
 \nonumber \\
&&   \quad +\bigg( \frac{-18+234v_Q^2+167v_Q^4-118v_Q^6}{18v_Q^2} + 
            \frac{-3-10v_Q^2+5v_Q^4}{2} \,\zeta_2 \bigg) \ln p  
 \nonumber \\ 
&&   \quad +\frac{-9+510v_Q^2-118v_Q^4}{9v_Q} + 
     v_Q \,(-27+5v_Q^2) \,\zeta_2 \, \Bigg]
     \,,
\label{eq::rccv}
\end{eqnarray}
and
\begin{equation}
\rho^R(M_Q^2,M_Q^2,s) = \frac{1}{3}\,
\int_{4M_Q^2/s}^{(1-2M_Q/\sqrt{s})^2}{\rm d}y\,
\int_{4M_Q^2/s}^{(1-\sqrt{y})^2}\frac{{\rm d}z}{z}\,
\left(1+\frac{2M_Q^2}{s z}\right)\sqrt{1-\frac{4M_Q^2}{s z}}
\,\,{\cal F}(y,z)\,,
\label{eq::rccr}
\end{equation}
with
\begin{eqnarray}
{\cal F}(y,z) & = &
\frac{\frac{8M_Q^4}{s^2} + \frac{4M_Q^2}{s}(1-y+z) - (1-y+z)^2 - 2(1+z)y}
 {1-y+z} \nonumber\\  
&&\times\ln\frac{1-y+z-\sqrt{1-\frac{4M_Q^2}{s y}}\,\Lambda^{1/2}(1,y,z)}
        {1-y+z+\sqrt{1-\frac{4M_Q^2}{s y}}\,\Lambda^{1/2}(1,y,z)}
\nonumber\\
&& -\sqrt{1-\frac{4M_Q^2}{s y}}\,\Lambda^{1/2}(1,y,z)\, 
 \left[ 1 + \frac{ \frac{16M_Q^4}{s^2}+\frac{8M_Q^2}{s} + 
        4\left(1+\frac{2M_Q^2}{s}\right)z }{
 (1-y+z)^2-\left(1-\frac{4M_Q^2}{s y}\right)\,\Lambda(1,y,z)} \, \right] \,,
\nonumber\\
\Lambda(1,y,z) &=& 1+y^2+z^2-2(y+z+y z)\,.
\end{eqnarray}
Note that $\rho^R(M_Q^2,M_Q^2,s)$ vanishes for $s\leq (4M_Q)^2$.

The numerical integration of \eqn{eq::rccr} would be a bottle neck of
\rhadf{}. But it turns out that, for $s>(4M_Q)^2$, $r_V^{\rm T}(s,M_Q)$
is extremely well approximated by the high energy expansion of
\reference{CHKSm12}, and it drastically shortens the running time of
\rhadf{}. Thus we use this expression as the default in \rhadf{} for
$s>(4M_Q)^2$, whereas for $s\leq (4M_Q)^2$ we can savely use
\eqn{eq::r2cc} since the double integral does not contribute in this
case.  For completeness, we list the expansion for $r_V^{\rm T}$ up to
$M_Q^{12}/s^6$~\cite{CHKSm12}:
\begin{equation}
\begin{split}
   r_V^{\rm T}(s,M_Q) &= \cf\tr\bigg\{
       - {11\over 8}
          + \zeta_3
          + {1\over 4} \lnsmu
       + {M_Q^2\over s}\, \bigg[
          - {13\over 2}
          + 3 \lnsmu
          \bigg]
\\&
       + \left({M_Q^2\over s}\right)^{2} \, \bigg[
            {2\over 3}
          + 18\,\zeta_2
          - 10\,\zeta_3
          + 12 \lnmsms
          - 3 \lnmsms^2
          + \left( {5\over 2}
          - 6 \lnmsms\right) \lnsmu
          \bigg]
\\&
       + \left({M_Q^2\over s}\right)^{3} \, \bigg[
            {4\over 3}
          + {352\over 9}\,\zeta_2
          + {350\over 9} \lnmsms
          - {76\over 9} \lnmsms^2
          + \left( - {188\over 27}
          - {116\over 9} \lnmsms\right) \lnsmu
          \bigg]
\\&
       + \left({M_Q^2\over s}\right)^{4} \, \bigg[
            {20233\over 288}
          + {533\over 6}\,\zeta_2
          + {1673\over 72} \lnmsms
          - {25\over 4} \lnmsms^2
\\&\mbox{\hspace{0.8cm}}
          + \left( - {983\over 36}
          - {203\over 6} \lnmsms\right) \lnsmu
          \bigg]
\\&
       + \left({M_Q^2\over s}\right)^{5} \, \bigg[
          - {54559\over 6750}
          + {3592\over 45}\,\zeta_2
          - {1157\over 75} \lnmsms
          + {4328\over 45} \lnmsms^2
\\&\mbox{\hspace{0.8cm}}
          + \left( - {61699\over 675}
          - {4676\over 45} \lnmsms\right) \lnsmu
          \bigg]
\\&
       + \left({M_Q^2\over s}\right)^{6} \, \bigg[
          - {9214697\over 6480}
          - 1105\,\zeta_2
          + {346981\over 540} \lnmsms
          + {18937\over 18} \lnmsms^2
\\&\mbox{\hspace{0.8cm}}
          + \left( - {84743\over 270}
          - {3064\over 9} \lnmsms\right) \lnsmu
          \bigg] \bigg\}+ \ldots\,,
\label{eq::rvctm12}
\end{split}
\end{equation}
where $\lnmsms = \ln(M_Q^2/s)$, $\lnsmu = \ln(s/\mu^2)$. The ellipse
denotes uncalculated higher order terms in $M_Q^2/s$.

The double-bubble contribution with zero outer mass and inner mass $M_2$ 
is given by~\cite{HJKT}
\begin{eqnarray}
  r_V^{\rm db}(s,0,M_2) = 
  \cf\tr\left[\varrho^V(M_2^2,s)
  +\theta(s-4M_2^2)
  \left(\varrho^R(M_2^2,s) +\frac{1}{4}\ln\frac{M_2^2}{\mu^2}\right)\right]\,,
  \label{eq::rvdbm2}
\end{eqnarray}
with
\begin{eqnarray}
\varrho^R(M_2^2,s)  & = &
 \frac{4}{3}\,\left( 1 - 6 x^2 \right) \,
   \left[\frac{1}{2}\,{\rm Li}_3({{1 - w}\over 2}) -
         \frac{1}{2}\,{\rm Li}_3({{1 + w}\over 2})
 \right.  \nonumber \\ & & \left.
\quad+\, {\rm Li}_3({{1 + w}\over {1 + a}}) -
      {\rm Li}_3({{1 - w}\over {1 - a}}) +
      {\rm Li}_3({{1 + w}\over {1 - a}}) -
      {\rm Li}_3({{1 - w}\over {1 + a}})
 \right.  \nonumber \\ & & \left.
\quad +\,\frac{1}{2}\,\ln ({{1 + w}\over {1 - w}})
  \left\{
   \zeta_2 -
   \frac{1}{12}\,\ln^2 (\frac{1+w}{1-w}) +
   \frac{1}{2}\,\ln^2(\frac{a-1}{a+1}) 
   \right.\right.\nonumber\\&&\left.\left.\mbox{}
   \qquad\qquad-\frac{1}{2}\,\ln (\frac{1+w}{2}) \ln (\frac{1-w}{2})
  \right\}
    \right]
    \nonumber\\ & &
 +\,\frac{1}{9}\,a\,\left( 19 + 46 x \right) \,
  \left(
   {\rm Li}_2({{1 + w}\over {1 + a}}) +
        {\rm Li}_2({{1 - w}\over {1 - a}}) -
        {\rm Li}_2({{1 + w}\over {1 - a}}) -
        {\rm Li}_2({{1 - w}\over {1 + a}}) \right.
    \nonumber\\ & &
  \left.
  \qquad\qquad+\,\ln ({{a - 1}\over {a + 1}})\,\ln ({{1 + w}\over {1 - w}})
    \right)
   \nonumber\\ & &
 +\,4 \left( {{19}\over {72}} + x + x^2  \right) \,
   \left(
    {\rm Li}_2(-\frac{1+w}{1-w}) - {\rm Li}_2(-\frac{1-w}{1+w})
    -\ln x\, \ln (\frac{1+w}{1-w}) \right)
  \nonumber\\ & &
 +\,7 \left( {{73}\over {189}} + {{74}\over {63}} x +
   x^2\right) \,\ln ({{1 + w}\over {1 - w}})
-\,\frac{1}{3}\left( {{2123}\over {108}} + \frac{2489}{54} x \right) \,w
   \,, 
\\
\varrho^V(M_2^2,s) & = &
 \frac{2}{3}\left(1-6x^2\right)\left({\rm Li}_3(A^2) - \zeta_3-
   2\zeta_2\ln A+\frac{2}{3}\ln^3 A\right)
   \nonumber\\& &
 +\,\frac{1}{9}\left(19+46x\right)\sqrt{1+4x}\left({\rm Li}_2(A^2)
   -\zeta_2+\ln^2 A\right)
   \nonumber\\& &
 +\,\frac{5}{36}\left(\frac{53}{3}+44x\right)\ln x+\frac{3355}{648}
  +\frac{119}{9}x
\,, 
\label{eq::rhov2}
\end{eqnarray}
where
\begin{eqnarray}
x=\frac{M_2^2}{s}\,,\quad a=\sqrt{1+4 x}\,, \quad w=\sqrt{1-4 x} \,,
\quad
A=\frac{\sqrt{1+4x}-1}{\sqrt{4 x}}\,.
\nonumber
\end{eqnarray}

In the limit $s\gg M_2^2$, the result for $r_V^{\rm db}(s,0,M_2)$ reads
\begin{eqnarray}
  r_V^{\rm db}(s,0,M_2) &=& \cf\tr\left(
    -\frac{11}{8}+\zeta_3+\frac{1}{4}\ln\frac{s}{\mu^2} 
    + {\cal O}\left(\frac{M_2^4}{s^2}\right)
  \right)
  \,.
  \label{eq::rvdb0m2}
\end{eqnarray}
Furthermore we need $r_V^{\rm db}(s,M_1,0)$ up to the quadratic mass terms
which enters the evaluation of $R_c^{(2)}$ in the limit
$s>s_{b}^{\rm thr}$. It is given by
\begin{eqnarray}
  r_V^{\rm db}(s,M_1,0) &=& \cf\tr\left(
    -\frac{11}{8}+\zeta_3+\frac{1}{4}\ln\frac{s}{\mu^2} 
    +\frac{M_1^2}{s} \left( -\frac{13}{2} + 3 \ln\frac{s}{\mu^2} \right)
    + {\cal O}\left(\frac{M_1^4}{s^2}\right)
  \right)
  \,.
  \nonumber\\
  \label{eq::rvdbm1}
\end{eqnarray}

In order to obtain the result for the double-bubble contribution in the 
limit $M_2^2\gg s$ one has to perform the decoupling of $M_2$ from the
coupling constant using Eq.~(\ref{eq::asdec}) 
and the one-loop approximation of Eq.~(\ref{eq:zetagOS}).
Taking into account the full $M_1$ dependence the result 
can be cast in the form~\cite{Seidiplom}
\begin{eqnarray}
  r_V^{\rm db,h}(s,M_1,M_2) &=& \cf\tr \frac{s}{45M_2^2}\Bigg[ 
  \left(
    -1 + \frac{12M_1^4}{s^2} + \frac{16M_1^6}{s^3}
  \right)
  \ln\frac{1 + \sqrt{1 - \frac{4M_1^2}{s}}}{1 - \sqrt{1 - \frac{4M_1^2}{s}}}
  \nonumber\\&&\mbox{}
  \qquad
  + 
  \left( \frac{22}{5} 
  + \frac{79M_1^2}{5s} + \frac{8M_1^4}{s^2}
  \right)\sqrt{1 - \frac{4M_1^2}{s}}
  \nonumber\\&&\mbox{}
  \qquad
  +
  \left(
  1 + \frac{2M_1^2}{s}
  \right)\ln\frac{M_2^2}{M_1^2} \sqrt{1 - \frac{4M_1^2}{s}}
  \,\Bigg]
  \,.
  \label{eq::rvdbts}
\end{eqnarray}
The expansion of $\cf\tr\varrho^V$ 
in the limit $M_2^2\gg s$ agrees with 
$r_V^{\rm db,h}$ for $M_1=0$ and reads
\begin{eqnarray}
  r_V^{\rm db,h}(s,0,M_2) &=& \cf\tr \frac{3s}{8M_2^2} 
  \left( \frac{176}{675} + \frac{8}{135}\ln\frac{M_2^2}{s} \right)
  \,.
\end{eqnarray}

The massless limits of the above expressions have been given in
\eqn{eq::2lml}.



\subsection{Analytical results for $R_Q^{(3)}(s)$}\label{app::3loop}
This section lists analytical expressions for the three-loop functions
defined in \sct{sec::3loop}. As already mentioned, only the high-energy
expansion is known, including quartic mass terms. We write these
functions as follows:
\begin{equation}
\begin{split}
  r_V^{(3)} &= r_0^{(3)} +
  \frac{M_Q^2}{s}\,r_{V,2}^{(3)} +
  \left(\frac{M_Q^2}{s}\right)^2\,
  r_{V,4}^{(3)}\,,
\qquad\qquad
\delta r_0^{(3)} = \frac{M_Q^2}{s}\,r_{0,2}^{(3)} +
  \left(\frac{M_Q^2}{s}\right)^2\,r_{0,4}^{(3)}\,.
\end{split}
\end{equation}
The results for the massless terms have been obtained in
\reference{3lm0,chet3l}, the quadratic mass terms in \reference{CKa3m2},
and the quartic terms in \reference{CHKa3m4}.  Adopting the on-shell
scheme for the mass $M_Q$, we get, for the contributions where the
massive quark couples to the external current:
\begin{equation}
\begin{split}
r_0^{(3)} &= {87029\over 288} 
      - {121\over 8}\,\zeta_2 
      - {1103\over 4}\,\zeta_3 
      + {275\over 6}\,\zeta_5
     + \lnsmu\,\Big(
          -{4321\over 48} 
          + {121\over 2}\,\zeta_3
          \Big) 
      + {121\over 16}\,\lnsmu^2 
\\&
      + n_f\,\bigg[
          -{7847\over 216} 
          + {11\over 6}\,\zeta_2 
          + {262\over 9}\,\zeta_3 
          - {25\over 9}\,\zeta_5
          + \lnsmu\,\Big(
              {785\over 72} 
              - {22\over 3}\,\zeta_3
              \Big) 
          - {11\over 12}\,\lnsmu^2 
          \bigg]
\\&
      + n_f^2\,\bigg[
          {151\over 162} 
          - {1\over 18}\,\zeta_2 
          - {19\over 27}\,\zeta_3
          + \lnsmu\,\Big(
              -{11\over 36} 
              + {2\over 9}\,\zeta_3
              \Big) 
          + {1\over 36}\,\lnsmu^2 
          \bigg] \,,
\end{split}
\end{equation}
\begin{equation}
\begin{split}
  r_{V,2}^{(3)} &=
       \frac{22351}{12} 
     - \frac{967}{2}\zeta_2
     - 16\zeta_2\ln2
     + \frac{502}{3}\zeta_3
     - \frac{5225}{6}\zeta_5
     + 378\lMs 
     - 9\lMs^2 
     \\&
     + \lnsmu\left( - \frac{2385}{4} - 132\lMs\right)
     + \frac{363}{4}\lnsmu^2 
     + n_f\bigg[
         -\frac{8429}{54} 
         + 42\zeta_2 
         - \frac{466}{27}\zeta_3 
         + \frac{1045}{27}\zeta_5
\\&\qquad
         - \frac{52}{3}\lMs
         + 2\lMs^2 
         + \lnsmu\left( \frac{389}{6} + 8\lMs\right)
         - 11\lnsmu^2
          \bigg]
\\&
     + n_f^2\left(\frac{125}{54} - \frac{2}{3}\zeta_2 - \frac{13}{9}\lnsmu
       + \frac{1}{3}\lnsmu^2
     \right)
    \,,
\end{split}
\end{equation}
\begin{equation}
\begin{split}
  r_{V,4}^{(3)} &=
        \frac{91015}{108} 
      - \frac{76}{9}\ln^4 2 
      + \frac{2564287}{540}\zeta_2
      - \frac{4568}{9}\zeta_2\ln 2
      \\&
      - \frac{128}{3}\zeta_2\ln^2 2
      + \frac{56257}{18}\zeta_3
      - \frac{1439}{3}\zeta_2\zeta_3
      - \frac{1565}{6}\zeta_4
      - \frac{3770}{3}\zeta_5
      - \frac{608}{3}a_4
      \\&
      + \lMs \bigg(-\frac{5536}{3}
            + 564\zeta_2
            - 24\zeta_2\ln 2
            + 416\zeta_3
            \bigg)
      - \frac{591}{4}\lMs^2
      + \frac{15}{2}\lMs^3
  \\&
      + \lnsmu\bigg(
      -\frac{5297}{12}
      - 1199\zeta_2 
      - 572\zeta_3 
      - 88\zeta_2\ln 2
      + \frac{4033}{4}\lMs 
      + \frac{165}{2}\lMs^2
      \bigg)
  \\&
      + \lnsmu^2\bigg(
      \frac{605}{8} 
      - \frac{363}{2}\lMs
      \bigg)
  \\&
      + n_f\bigg[
            - \frac{21011}{216} 
            + \frac{8}{27}\ln^4 2
            - \frac{3544}{9}\zeta_2
            - \frac{176}{9}\zeta_2\ln 2
            + \frac{32}{9}\zeta_2\ln^2 2
            - \frac{2323}{9}\zeta_3
\\&\quad
            + \frac{700}{9}\zeta_4
            + \frac{440}{9}\zeta_5
            + \frac{64}{9} a_4
            + \lMs\bigg(
                  \frac{2419}{12} 
                + \frac{44}{3}\zeta_2
                + \frac{16}{3}\zeta_2 \ln2
                + \frac{28}{3}\zeta_3
                \bigg) 
\\&\quad
            - \frac{157}{6}\lMs^2
            - \frac{2}{3}\lMs^3
            + \lnsmu\bigg(
            \frac{1879}{36}
            + \frac{416}{3}\zeta_2
            + \frac{148}{3}\zeta_3
            + \frac{16}{3}\zeta_2\ln 2
\\&\qquad
            - \frac{361}{3}\lMs
            + 6\lMs^2 
            \bigg)
            + \lnsmu^2\bigg(
            -\frac{55}{6}
            + 22\lMs
            \bigg)
            \bigg]
\\&
      + n_f^2\bigg[
             \frac{35}{18} 
           + \frac{25}{3}\zeta_2
           + \frac{112}{27}\zeta_3
           + \lMs\bigg(
                 - \frac{94}{27} 
                 - \frac{8}{3}\zeta_2
                 \bigg)
           + \frac{13}{9}\lMs^2
           - \frac{2}{9}\lMs^3
\\&\quad
           + \lnsmu\bigg(
           -\frac{35}{27}
           - 4\zeta_2 
           - \frac{8}{9}\zeta_3
           + 3\lMs 
           - \frac{2}{3}\lMs^2
           \bigg)
           + \lnsmu^2\bigg(
           \frac{5}{18}
           - \frac{2}{3}\lMs
           \bigg)
           \bigg]\,,
\end{split}
\end{equation}
where $n_f$ is the number of active flavors,
$a_4=\mbox{Li}_4(1/2)\approx 0.517479$, 
$\zeta_2 = \pi^2/6\approx 1.64492$,
$\zeta_3 \approx 1.20206$,
$\zeta_4 = \pi^4/90 \approx 1.08232$,
$\zeta_5 \approx 1.03693$,
$\lMs=\ln (M_Q^2/s)$ and $\lnsmu = \ln (s/\mu^2)$.
The mass corrections in the case where the massive quark only appears as
insertion in gluon lines reads:
\begin{equation}
\begin{split}
r_{0,2}^{(3)} &=
-80 
+ 60\,\zeta_3
+ n_f\,\bigg(
{32\over 9} 
- {8\over 3}\,\zeta_3
\bigg) \,,
\end{split}
\end{equation}
\begin{equation}
\begin{split}
   r_{0,4}^{(3)} &=
       - \frac{4217}{48} 
       + 15\zeta_2
       + \frac{139}{3}\zeta_3
       + \frac{50}{3}\zeta_5
       + \lMs\left(\frac{97}{4} - 38\zeta_3\right)
       - 2\lMs^2 
       \\&
       + \lnsmu\bigg(
       - \frac{143}{6}
       + 22\zeta_3
       + \frac{11}{2}\lMs 
       \bigg)
       \\&
       + n_f\bigg[
             \frac{457}{108} 
           - \frac{2}{3}\zeta_2
           - \frac{22}{9}\zeta_3
           + \lMs\left(
                  - \frac{13}{18} 
                  + \frac{4}{3}\zeta_3
                 \right)
                 + \lnsmu\bigg(
                 \frac{13}{9}
                 - \frac{1}{3}\lMs 
                 - \frac{4}{3}\zeta_3
                 \bigg)
                 \bigg]
  \,.
\end{split}
\end{equation}
%


\subsection{Analytical results for $R_Q^{(4)}(s)$}\label{app::4loop}

With the help of the renormalization group equation the logarithmic
contributions of the massless order $\alpha_s^4$ result can be reconstructed
exactly. Thus, for $\mu^2\not= s$ Eqs.~(\ref{eq::r04}) and~(\ref{eq::rv4m2}) 
take the form
\begin{equation}
  \begin{split}
    \bar r_0^{(4),{\rm 0L}}(s) &= -186
    +\left(-\frac{520175}{192} 
      + \frac{3993}{32}\zeta_2 
      + \frac{38643}{16}\zeta_3
      - \frac{3025}{8}\zeta_5
    \right)\lnsmu
\\&\qquad
    +\left(
        \frac{49775}{128} 
      - \frac{3993}{16}\zeta_3
    \right)\lnsmu^2
    -\frac{1331}{64}\lnsmu^3
    \,,\\
    \bar r_0^{(4),{\rm 1L}}(s) &= 21.3
    +\left(
      \frac{188521}{384}
      - \frac{363}{16}\zeta_2
      - \frac{9695}{24}\zeta_3
      + \frac{275}{6}\zeta_5
    \right)\lnsmu
\\&\qquad
    +\left(
      -\frac{2263}{32} 
      + \frac{363}{8}\zeta_3
    \right)\lnsmu^2
    +\frac{121}{32}\lnsmu^3
    \,,\\
    r_0^{(4),{\rm 2L}}(s) &= 
\frac{1045381}{15552} 
-\frac{593}{432}  \pi^2
-\frac{40655}{864}  \,\zeta_{3}
+ \frac{11}{12}  \pi^2 \,\zeta_{3}
+\frac{5}{6}  \,\zeta_3^2
-\frac{260}{27}  \,\zeta_{5}
\\&
    +\left(
      -\frac{94693}{3456}
      + \frac{11}{8}\zeta_2 
      + \frac{257}{12}\zeta_3
      - \frac{25}{18}\zeta_5
    \right)\lnsmu
    +\left(
      \frac{593}{144} 
      - \frac{11}{4}\zeta_3
    \right)\lnsmu^2
    -\frac{11}{48}\lnsmu^3
    \,,\\
    r_0^{(4),{\rm 3L}}(s) &= 
-\frac{6131}{5832} 
+\frac{11}{432}  \pi^2
+\frac{203}{324}  \,\zeta_{3}
-\frac{1}{54}  \pi^2 \,\zeta_{3}
+\frac{5}{18}  \,\zeta_{5}
\\&
    +\left(
      \frac{151}{324}
      - \frac{1}{36}\zeta_2
      - \frac{19}{54}\zeta_3
    \right)\lnsmu
    +\left(
      -\frac{11}{144}
      + \frac{1}{18}\zeta_3
    \right)\lnsmu^2
    +\frac{1}{216}\lnsmu^3
    \,,\\
    \bar{r}_{V,2}^{(4),{\rm 0L}}(s) &=
    7110
    - \frac{9855433}{864} 
    - \frac{535211}{270}\zeta_2
    + \frac{3194}{9}\zeta_2\ln 2
    + \frac{128}{3} \zeta_2\ln^2 2
    + \frac{76}{9}\ln^4 2
    - \frac{9655}{18}\zeta_3
    \\&
    + \frac{1439}{3}\zeta_2 \zeta_3
    + \frac{1565}{6}\zeta_4
    + \frac{18925}{9}\zeta_5
    + \frac{608}{3} a_4
    \\&
    + \left(
      \frac{178849}{24} 
      - 659 \zeta_2
      + 56 \zeta_2\ln2
      + \frac{938}{3}\zeta_3
      - \frac{5225}{3}\zeta_5
    \right)\lMs 
    - \frac{2313}{8}\lMs^2
    + \frac{21}{2}\lMs^3
    \\&
    + \left(
      - \frac{538849}{32} 
      - \frac{6849}{2}\lMs
      + \frac{297}{4}\lMs^2
      + \frac{31911}{8} \zeta_2
      + 132\zeta_2\ln2
      - \frac{2761}{2}\zeta_3
    \right.\\&\left.
      + \frac{57475}{8}\zeta_5
    \right)\lnsmu 
    + \left(\frac{85437}{32} 
      + \frac{1089}{2}\lMs
    \right)\lnsmu^2 
    - \frac{3993}{16} \lnsmu^3
    \,,\\
    \bar{r}_{V,2}^{(4),{\rm 1L}}(s) &=
    - 1430
    + \frac{604967}{648} 
    + \frac{689}{3}\zeta_2
    + \frac{76}{3}\zeta_2\ln2
    - \frac{32}{9}\zeta_2\ln^2 2
    - \frac{8}{27}\ln^4 2
    + \frac{10622}{81}\zeta_3
    \\&
    - \frac{610}{9}\zeta_4
    - \frac{8360}{81}\zeta_5
    - \frac{64}{9} a_4
    + \left(
      - \frac{35899}{54} 
      + \frac{64}{3}\zeta_2
      - \frac{16}{3}\zeta_2\ln2
      - \frac{1436}{27}\zeta_3
    \right.\\&\left.
      + \frac{2090}{27}\zeta_5
    \right)\lMs
    + \frac{133}{2}\lMs^2
    - \frac{10}{3}\lMs^3
    + \left(
      \frac{713191}{288} 
      - \frac{2353}{4}\zeta_2
      - 8\zeta_2 \ln2
      + \frac{4069}{18}\zeta_3
    \right.\\&\left.
      - \frac{13585}{18}\zeta_5
      + 370\lMs 
      - 21\lMs^2 
    \right)\lnsmu
    + \left(-\frac{3643}{8} - 66\lMs\right)\lnsmu^2 
    + \frac{363}{8}\lnsmu^3
    \,,\\
    r_{V,2}^{(4),{\rm 2L}}(s) &=
    \frac{114253}{648} 
    - \frac{415}{6}\zeta_2
    - \frac{995}{324}\zeta_3
    + \frac{53}{9}\zeta_3^2
    - \frac{3610}{81}\zeta_5 
    + \left(
      \frac{199}{18} 
      + \frac{4}{3}\zeta_2
    \right) \lMs
    - \frac{13}{6}\lMs^2
    \\&
    + \frac{2}{9}\lMs^3
    + \left(
      - \frac{90835}{864} 
      + \frac{53}{2}\zeta_2
      - \frac{233}{27}\zeta_3
      + \frac{1045}{54}\zeta_5
      - \frac{26}{3}\lMs
      + \lMs^2 
    \right)\lnsmu
    \\&
    + \left(\frac{95}{4} + 2\lMs\right)\lnsmu^2 
    - \frac{11}{4}\lnsmu^3
    \,,\\
    r_{V,2}^{(4),{\rm 3L}}(s) &=
    - \frac{2705}{1944} 
    + \frac{13}{18}\zeta_2
    + \left(\frac{125}{108} - \frac{1}{3}\zeta_2\right)\lnsmu 
    - \frac{13}{36}\lnsmu^2
    + \frac{1}{18}\lnsmu^3
    \,,
\end{split}
\label{eq::r04log}
\end{equation}
where the first number in $\bar r_{0}^{(4),{\rm 0L}}$, $\bar
r_{0}^{(4),{\rm 1L}}$, $\bar r_{V,2}^{(4),{\rm 0L}}$, and $\bar
r_{V,2}^{(4),{\rm 1L}}$ corresponds to the estimates obtained with the
help of the results of Refs.\,\cite{BCK,BCKRadcor02}.


\subsection{\label{app::renorm}Renormalization group coefficients}
This section collects the formulas used for the evaluation of the
running coupling constant and the $\msbar$ quark masses.
It also contains the conversion formulas for transforming $\msbar$ quark
masses to their on-shell values.

The coefficients of the $\beta$-function defined in \eqn{eq::beta}
read~\cite{betafun4}:
\begin{eqnarray}
\renewcommand{\arraystretch}{ 1.3}
 \beta_0^{(n_f)} & = & {1\over 4}\left[11 - \frac{2}{3} n_f \right]\,,  \nonumber\\
\beta_1^{(n_f)} & = & {1\over 16}\left[ 102 - \frac{38}{3} n_f
 \right]\,, \nonumber \\
 \beta_2^{(n_f)} & = & {1\over 64}\left[ \frac{2857}{2} - \frac{5033}{18} n_f +
 \frac{325}{54} n_f^2 \right]\,, \nonumber \\
 \beta_3^{(n_f)} & = &  {1\over 256} \bigg[
 \left( \frac{149753}{6} + 3564 \zeta_3 \right)
        - \left( \frac{1078361}{162} + \frac{6508}{27} \zeta_3 \right) n_f
  \nonumber \\ & &
       + \left( \frac{50065}{162} + \frac{6472}{81} \zeta_3 \right) n_f^2
       +  \frac{1093}{729}  n_f^3
       \bigg]\,.
\end{eqnarray}

The coefficients of the $\gamma_m$-function of \eqn{eq::gamma}
are~\cite{gammam4}:
\begin{eqnarray}
\renewcommand{\arraystretch}{ 1.3}
\label{eq:gamma}
\gamma_{m,0}^{(n_f)} & = & 1 \,,  \nonumber\\
\gamma_{m,1}^{(n_f)} & = & \frac{1}{16}\left[ \frac{202}{3}
                 - \frac{20}{9} n_f \right] \,,  
  \nonumber \\
\gamma_{m,2}^{(n_f)} & = & \frac{1}{64} \left[1249+\left( - \frac{2216}{27} 
          - \frac{160}{3}\zeta_3 \right)n_f 
               - \frac{140}{81} n_f^2 \right]\,,
\nonumber \\
 \gamma_{m,3}^{(n_f)} & = & \frac{1}{256} \left[ 
       \frac{4603055}{162} + \frac{135680}{27}\zeta_3 - 8800\zeta_5
       \right.
\nonumber\\&&
 +\left(- \frac{91723}{27} - \frac{34192}{9}\zeta_3 
    + 880\zeta_4 + \frac{18400}{9}\zeta_5 \right) n_f
 \nonumber \\ & & \left.
 +\left( \frac{5242}{243} + \frac{800}{9}\zeta_3 
    - \frac{160}{3}\zeta_4 \right) n_f^2
 +\left(- \frac{332}{243} + \frac{64}{27}\zeta_3 \right) n_f^3 \right]\,.
\end{eqnarray}

The decoupling coefficient for $\alpha_s$ of \eqn{eq::asdec} 
is~\cite{Chetyrkin:1997un}:
\begin{eqnarray}
  \left(\zeta_g\right)^2 &=& 1 +
  \frac{\alpha_s^{(n_f)}(\mu)}{\pi}
  \left(
    -\frac{1}{6}\ln\frac{\mu^2}{M_h^2}
  \right)
  +\left(\frac{\alpha_s^{(n_f)}(\mu)}{\pi}\right)^2
  \left(
    -\frac{7}{24} 
    -\frac{19}{24}\ln\frac{\mu^2}{M_h^2}
    +\frac{1}{36}\ln^2\frac{\mu^2}{M_h^2}
  \right)
  \nonumber\\
  &&{}+\left(\frac{\alpha_s^{(n_f)}(\mu)}{\pi}\right)^3
  \left[
    -\frac{58933}{124416}
    -\frac{2}{3}\zeta_2\left(1+\frac{1}{3}\ln2\right)
    -\frac{80507}{27648}\zeta_3
    -\frac{8521}{1728}\ln\frac{\mu^2}{M_h^2}
  \right.\nonumber\\
  &&{}-\left.
    \frac{131}{576}\ln^2\frac{\mu^2}{M_h^2}
    -\frac{1}{216}\ln^3\frac{\mu^2}{M_h^2} 
    +n_l\left(
      \frac{2479}{31104}
      +\frac{\zeta_2}{9}
      +\frac{409}{1728}\ln\frac{\mu^2}{M_h^2} 
    \right)
  \right]
  \,,
  \label{eq:zetagOS}
\end{eqnarray}
and for the quark masses, \eqn{eq::mqdec}
\cite{Chetyrkin:1997un}:
\begin{eqnarray}
\zeta_m &=&   1
  +\left(\frac{\alpha_s^{(n_f)}(\mu)}{\pi}\right)^2
  \left(\frac{89}{432} 
    -\frac{5}{36}\ln\frac{\mu^2}{M_h^2}
    +\frac{1}{12}\ln^2\frac{\mu^2}{M_h^2}
  \right)
  +\left(\frac{\alpha_s^{(n_f)}(\mu)}{\pi}\right)^3
  \left[
    \frac{1871}{2916} 
  \right.
  \nonumber\\
  &&
  \left.\mbox{}
    - \frac{407}{864}\zeta_3
    +\frac{5}{4}\zeta_4
    - \frac{1}{36}B_4
    +\left(\frac{121}{2592}
      - \frac{5}{6}\zeta_3\right)\ln\frac{\mu^2}{M_h^2}
    + \frac{319}{432}\ln^2\frac{\mu^2}{M_h^2}
  \right.
  \nonumber\\
  &&
  \left.\mbox{}
    + \frac{29}{216}\ln^3\frac{\mu^2}{M_h^2}
    +n_l\left(
      \frac{1327}{11664}
      - \frac{2}{27}\zeta_3
      - \frac{53}{432}\ln\frac{\mu^2}{M_h^2}
      - \frac{1}{108}\ln^3\frac{\mu^2}{M_h^2}
    \right)
  \right]
  \,,
  \label{eq:zetamOS} 
\end{eqnarray}
where $B_4 \approx -1.76280$,  $n_l=n_f-1$ is the number of light
(massless) quarks and $M_h$ is the pole mass of the heavy quark. The
versions of Eqs.~(\ref{eq:zetagOS}) and~(\ref{eq:zetamOS}) where the
$\msbar$ mass is used as parameter can be found in
Refs.\,\cite{Chetyrkin:1997sg,Chetyrkin:1997un,Chetyrkin:2000yt,Steinhauser:2002rq}.

The conversion from $\msbar$ to on-shell quark masses 
reads~\cite{GraBroGraSch90,CheSte99,MelRit99}:
\begin{eqnarray}
  \frac{M_Q}{\mqbar{Q}(\mu)} &=&
  1 
  + \frac{\alpha_s^{(n_f)}(\mu)}{\pi} 
    \left( \frac{4}{3} + \lmumms \right)
  + \left(\frac{\alpha_s^{(n_f)}(\mu)}{\pi}\right)^2
  \Bigg[
  \frac{307}{32}
  + \left(2+\frac{2}{3}\ln2 \right)\zeta_2 
  - \frac{1}{6}\zeta_3
  \nonumber\\&&\mbox{}
  + \frac{493}{72} \lmumms
  + \frac{43}{24}\lmumms^2
  + n_l\left(
    - \frac{71}{144} 
    - \frac{1}{3}\zeta_2
    - \frac{13}{36} \lmumms - \frac{1}{12} \lmumms^2
  \right)
  \Bigg]
  \nonumber\\&&\mbox{}
  + \left(\frac{\alpha_s^{(n_f)}(\mu)}{\pi}\right)^3
  \Bigg[
  \frac{8481925}{93312} 
+ \zeta_2\left(\frac{652841}{6480} - \frac{1439}{72}\zeta_3\right)
+ \frac{58}{27}\zeta_3
- \frac{3475}{432}\zeta_4
  \nonumber\\&&\mbox{}
+ \frac{1975}{216}\zeta_5
- \frac{220}{27}\,a_4
- \frac{575}{27}\zeta_2\ln 2
- \frac{44}{27}\zeta_2\ln^2 2
- \frac{55}{162}\ln^4 2
  \nonumber\\&&\mbox{}
+ \lmumms\left(
          \frac{177305}{2592}
         +\frac{37}{3}\zeta_2 
         -\frac{67}{36}\zeta_3 
         +\frac{37}{9}\zeta_2\ln 2
         \right)
+ \frac{19315}{864}\lmumms^2
+ \frac{1591}{432}\lmumms^3
  \nonumber\\&&\mbox{}
+ n_l\Bigg(
     - \frac{231847}{23328} 
     - \frac{991}{108}\zeta_2
     - \frac{241}{72}\zeta_3
     + \frac{305}{108}\zeta_4
     + \frac{8}{27}\,a_4
     - \frac{22}{27}\zeta_2\ln 2
  \nonumber\\&&\mbox{}
     + \frac{4}{27}\zeta_2 \ln^2 2
     + \frac{1}{81} \ln^4 2
     + \lmumms \left(
          -\frac{10129}{1296} 
          -\frac{49}{18}\zeta_2
          -\frac{7}{9}\zeta_3
          -\frac{2}{9}\zeta_2\ln 2
              \right)
  \nonumber\\&&\mbox{}
     - \frac{1103}{432}\lmumms^2
     - \frac{10}{27}\lmumms^3
     \Bigg)
+ n_l^2\Bigg(
         \frac{2353}{23328} 
       + \frac{13}{54}\zeta_2
       + \frac{7}{54}\zeta_3
       + \lmumms\left(\frac{89}{648} + \frac{1}{9}\zeta_2\right)
  \nonumber\\&&\mbox{}
       + \frac{13}{216}\lmumms^2
       + \frac{1}{108} \lmumms^3
       \Bigg) 
  \Bigg]
  \,,
  \label{eq:zminv}
\end{eqnarray}
where $a_4 = {\rm Li}_4(1/2)\approx 0.517479$, $\lmumms =
\ln(\mu^2/\mqbar{Q}^2(\mu))$, and $n_l=n_f-1$.  Other variants of
Eq.~(\ref{eq:zminv}) are listed in~\reference{Chetyrkin:2000yt}.


\end{appendix}


\def\jref#1#2#3{{\bf #1} (#3) #2}
\def\app#1#2#3{{\it Act.~Phys.~Pol.~}\jref{\bf B #1}{#2}{#3}}
\def\apa#1#2#3{{\it Act.~Phys.~Austr.~}\jref{\bf#1}{#2}{#3}}
\def\annphys#1#2#3{{\it Ann.~Phys.~}\jref{\bf #1}{#2}{#3}}
\def\cmp#1#2#3{{\it Comm.~Math.~Phys.~}\jref{\bf #1}{#2}{#3}}
\def\cpc#1#2#3{{\it Comp.~Phys.~Commun.~}\jref{\bf #1}{#2}{#3}}
\def\epjc#1#2#3{{\it Eur.\ Phys.\ J.\ }\jref{\bf C #1}{#2}{#3}}
\def\fortp#1#2#3{{\it Fortschr.~Phys.~}\jref{\bf#1}{#2}{#3}}
\def\ijmpc#1#2#3{{\it Int.~J.~Mod.~Phys.~}\jref{\bf C #1}{#2}{#3}}
\def\ijmpa#1#2#3{{\it Int.~J.~Mod.~Phys.~}\jref{\bf A #1}{#2}{#3}}
\def\jcp#1#2#3{{\it J.~Comp.~Phys.~}\jref{\bf #1}{#2}{#3}}
\def\jetp#1#2#3{{\it JETP~Lett.~}\jref{\bf #1}{#2}{#3}}
\def\jhep#1#2#3{{\it J.~High~Energy~Phys.~}\jref{\bf #1}{#2}{#3}}
\def\mpl#1#2#3{{\it Mod.~Phys.~Lett.~}\jref{\bf A #1}{#2}{#3}}
\def\nima#1#2#3{{\it Nucl.~Inst.~Meth.~}\jref{\bf A #1}{#2}{#3}}
\def\npb#1#2#3{{\it Nucl.~Phys.~}\jref{\bf B #1}{#2}{#3}}
\def\nca#1#2#3{{\it Nuovo~Cim.~}\jref{\bf #1A}{#2}{#3}}
\def\plb#1#2#3{{\it Phys.~Lett.~}\jref{\bf B #1}{#2}{#3}}
\def\prc#1#2#3{{\it Phys.~Reports }\jref{\bf #1}{#2}{#3}}
\def\prd#1#2#3{{\it Phys.~Rev.~}\jref{\bf D #1}{#2}{#3}}
\def\pR#1#2#3{{\it Phys.~Rev.~}\jref{\bf #1}{#2}{#3}}
\def\prl#1#2#3{{\it Phys.~Rev.~Lett.~}\jref{\bf #1}{#2}{#3}}
\def\pr#1#2#3{{\it Phys.~Reports }\jref{\bf #1}{#2}{#3}}
\def\ptp#1#2#3{{\it Prog.~Theor.~Phys.~}\jref{\bf #1}{#2}{#3}}
\def\ppnp#1#2#3{{\it Prog.~Part.~Nucl.~Phys.~}\jref{\bf #1}{#2}{#3}}
\def\sovnp#1#2#3{{\it Sov.~J.~Nucl.~Phys.~}\jref{\bf #1}{#2}{#3}}
\def\sovus#1#2#3{{\it Sov.~Phys.~Usp.~}\jref{\bf #1}{#2}{#3}}
\def\tmf#1#2#3{{\it Teor.~Mat.~Fiz.~}\jref{\bf #1}{#2}{#3}}
\def\tmp#1#2#3{{\it Theor.~Math.~Phys.~}\jref{\bf #1}{#2}{#3}}
\def\yadfiz#1#2#3{{\it Yad.~Fiz.~}\jref{\bf #1}{#2}{#3}}
\def\zpc#1#2#3{{\it Z.~Phys.~}\jref{\bf C #1}{#2}{#3}}
\def\ibid#1#2#3{{ibid.~}\jref{\bf #1}{#2}{#3}}


\end{document}